\documentclass[iop]{emulateapj}

\usepackage{amssymb,amsmath,amsfonts,amsbsy}
\usepackage{color}
\usepackage{natbib}
\usepackage{enumerate}
\usepackage{graphicx}

\begin{document}

\title{Cosmological Parameter Estimation Using the Genus Amplitude - Application to Mock Galaxy Catalogs}

\author{Stephen Appleby$^{a}$}\email{stephen@kias.re.kr}
\author{Changbom Park$^{a}$}
\author{Sungwook E. Hong$^{b}$}
\author{Juhan Kim$^{c}$}
\affiliation{$^{a}$School of Physics, Korea Institute for Advanced Study, 85
Hoegiro, Dongdaemun-gu, Seoul 02455, Korea}
\affiliation{$^{b}$Korea Astronomy and Space Science Institute, 776 Daedeokdae-ro, Yuseong-gu, Daejeon 34055, Korea}
\affiliation{$^{c}$Center for Advanced Computation, Korea Institute for Advanced Study, 85 Hoegiro, Dongdaemun-gu, Seoul 02455, Korea}

\begin{abstract}
We study the topology of the matter density field in two dimensional slices, and consider how we can use the amplitude $A$ of the genus for cosmological parameter estimation. Using the latest Horizon Run 4 simulation data, we calculate the genus of the smoothed density field constructed from lightcone mock galaxy catalogs. Information can be extracted from the amplitude of the genus by considering both its redshift evolution and magnitude. The constancy of the genus amplitude with redshift can be used as a standard population, from which we derive constraints on the equation of state of dark energy $w_{\rm de}$ - by measuring $A$ at $z \sim 0.1$ and $z \sim 1$, we can place an order $\Delta w_{\rm de} \sim {\cal O}(15\%)$ constraint on $w_{\rm de}$. By comparing $A$ to its Gaussian expectation value we can potentially derive an additional stringent constraint on the matter density $\Delta \Omega_{\rm mat} \sim 0.01$. We discuss the primary sources of contamination associated with the two measurements - redshift space distortion and shot noise. With accurate knowledge of galaxy bias, we can successfully remove the effect of redshift space distortion, and the combined effect of shot noise and non-linear gravitational evolution is suppressed by smoothing over suitably large scales $R_{\rm G} \ge 15 {\rm Mpc}/h$. Without knowledge of the bias, we discuss how joint measurements of the two and three dimensional genus can be used to constrain the growth factor $\beta = f/b$. The method can be applied optimally to redshift slices of a galaxy distribution generated using the drop-off technique. 
\end{abstract}

\maketitle

\section{\label{sec:in}Introduction}

The observed distribution of galaxies provides information regarding the initial conditions, composition and evolution of the Universe. Constructing statistics from which we can extract information remains an open field of research. Standard approaches involving the two point statistics have provided impressive constraints on cosmological parameters \citep{Park:1994fa, doi:10.1111/j.1365-2966.2011.19077.x, Eisenstein:2005su,Cole:2005sx,Anderson:2012sa}, however, the power spectrum alone cannot be used to probe the phase correlations of the field. One could generalise and study the N-point statistics, but these quantities become increasingly difficult to measure. This fact, coupled to other complications such as redshift space distortions and bias, has led to the construction of alternative statistics which are sensitive to different properties of the field. 

One such example are the Minkowski Functionals, a set of scalar quantities that encode the morphological and topological properties of an excursion set of the underlying density field. When measuring these quantities using the density field reconstructed from galaxy data (or any tracer of the underlying dark matter), we are effectively measuring ratios of cumulants of the field. For a Gaussian field, all information is contained in the power spectrum and in this case the shape of the genus curve is completely fixed, with only its amplitude varying as a function of smoothing scale and power spectrum shape. For a non-Gaussian field such as dark matter in the late Universe, the shape of the genus curve also carries information beyond the two point statistics. The Minkowski Functionals have been studied within the context of cosmology for nearly three decades \citep{Gott:1986uz,1989ApJ...340..625G,Ryden:1988rk,1989ApJ...345..618M,1992ApJ...387....1P,1991ApJ...378..457P,Matsubara:1994we,1996ApJ...457...13M,Schmalzing:1995qn,2005ApJ...633....1P}
. In this paper we focus predominantly on the two dimensional genus, which has been studied extensively in the literature \citep{1989ApJ...345..618M,1991MNRAS.250...75C,1992ApJ...387....1P,1993MNRAS.260..572C,Colley:1999rn, 2001ApJ...553...33P, 2002ApJ...570...44H,1997ApJ...489..471C, Gott:2006za,2015ApJ...814....6W}, and has been applied to both the CMB \citep{1990ApJ...352....1G,Schmalzing:1997uc,Hikage:2006fe,Ducout:2012it} and large scale structure \citep{1989ApJ...345..618M,1992ApJ...387....1P,1992ApJ...385...26G,Park:2009ja,Zunckel:2010eh,Speare:2013qma,Blake:2013noa,Matsubara:2000dg,Hikage:2006fe,Wang:2010ug,James:2011wm,0004-637X-489-2-457,Colley:1999rn, Watts:2017lzm, 1991MNRAS.250...75C,Hikage:2002ki,Choi:2010sx}. 

Galaxy data is typically provided as redshifts and angular positions on the sky. If we use this information to generate redshift shells and then calculate the angular power spectrum of galaxies within that shell, we have not made any inference regarding cosmology (other than statistical isotropy and homogeneity). However, the requirement of a cosmological model enters in a number of ways. One is when we wish to use the observed power spectrum (which is measured in redshift space) to infer real space quantities. A second is that when comparing the angular power spectrum (or genus) in different redshift slices we must account for the fact that different redshift shells occupy different surface areas. Finally, if use a constant physical smoothing scale $R_{\rm G}$ when smoothing the two dimensional density shells at different redshifts then we must relate angular $\theta_{\rm G}$ and physical $R_{\rm G}$ scales. All of the above requirements force us to adopt a distance-redshift relation. 

The dependence of the genus amplitude on redshift has been proposed as a test of cosmology by \citet{Park:2009ja} (see also \citep{Zunckel:2010eh,Speare:2013qma}). If we smooth the galaxy density field on large scales, then the topology is a conserved quantity. This means that if we measure the genus at successive redshifts, the amplitude of the genus should be constant on condition that we choose the correct cosmological model to measure distances. If we select incorrect cosmological parameters, then the physical scale at which we smooth the field and the number of structures in a unit volume will systematically evolve from low to high redshift. Hence for a given smoothing scale $R_{\rm G}$, the aim is to find the cosmology which minimizes the evolution of genus amplitude with redshift. 

Additional constraining power can be utilised if we smooth the field over sufficiently large scales - in this case we can directly compare the measured genus amplitude to its Gaussian expectation value. To leading order the amplitude is unaffected by non-linear gravitational collapse. This fact allows us to use this quantity to study the underlying linear power spectrum of the dark matter field. Thus there are two distinct approaches that we can use to extract information - using the amplitude of the genus and its redshift evolution.

In this work we will consider both approaches and how one can use them to impose cosmological parameter constraints. In the following section we test the sensitivity of the genus amplitude to cosmological parameters, adopting a Gaussian field. We reproduce the expected signal in a mock Gaussian density field, and then apply the same analysis to mock galaxy lightcone data. We discuss the two principle sources of contamination to the signal in section \ref{sec:sys} - redshift space distortion and shot noise. Correcting for these `systematics', we generate projected parameter constraints using both methods in section \ref{sec:ampr}. Finally, in section \ref{sec:amp} we consider the possibility of jointly using two and three dimensional topological information to simultaneously constrain $\Omega_{\rm mat}$ and the linear growth factor $\beta = f/b$.

\section{Genus of Two and Three Dimensional Density Fields}
\label{sec:1}

We begin with a discussion of Gaussian random fields. The low redshift dark matter density field that we wish to study is three dimensional, but from it we can generate two dimensional subsets by taking slices of width $\Delta$ along the line of sight (or in principle any direction). We label the full density field $\delta_{\rm 3D}$ and the two dimensional subsets $\delta_{\rm 2D}$. The two dimensional power spectrum $P_{\rm 2D}$ of $\delta_{\rm 2D}$ can be expressed in terms of its three dimensional counterpart as  

\begin{equation} P_{\rm 2D}(k_{\perp}) = {2 \over \pi} \int dk_{3} P_{\rm 3D}(\sqrt{\vec{k}^{2}_{\perp}+k^{2}_{\rm 3}}) {\sin^{2} [k_{3} \Delta] \over k_{3}^{2} \Delta^{2}}   \end{equation}

\noindent where $P_{\rm 3D}$ is the power spectrum of the three dimensional field $\delta_{\rm 3D}$, and we have performed real space top hat smoothing along an arbitrary $x_{3}$ direction, where $\Delta$ is the thickness of the slice (in units of ${\rm Mpc}/h$). We fix the line of sight to be parallel to an arbitrary axis $x_{3}$ in what follows. $\vec{k}_{\perp}=(k_{1},k_{2})$ and $k_{3}$ are the Fourier modes parallel and perpendicular to the plane.

\begin{figure}
  \includegraphics[width=0.45\textwidth]{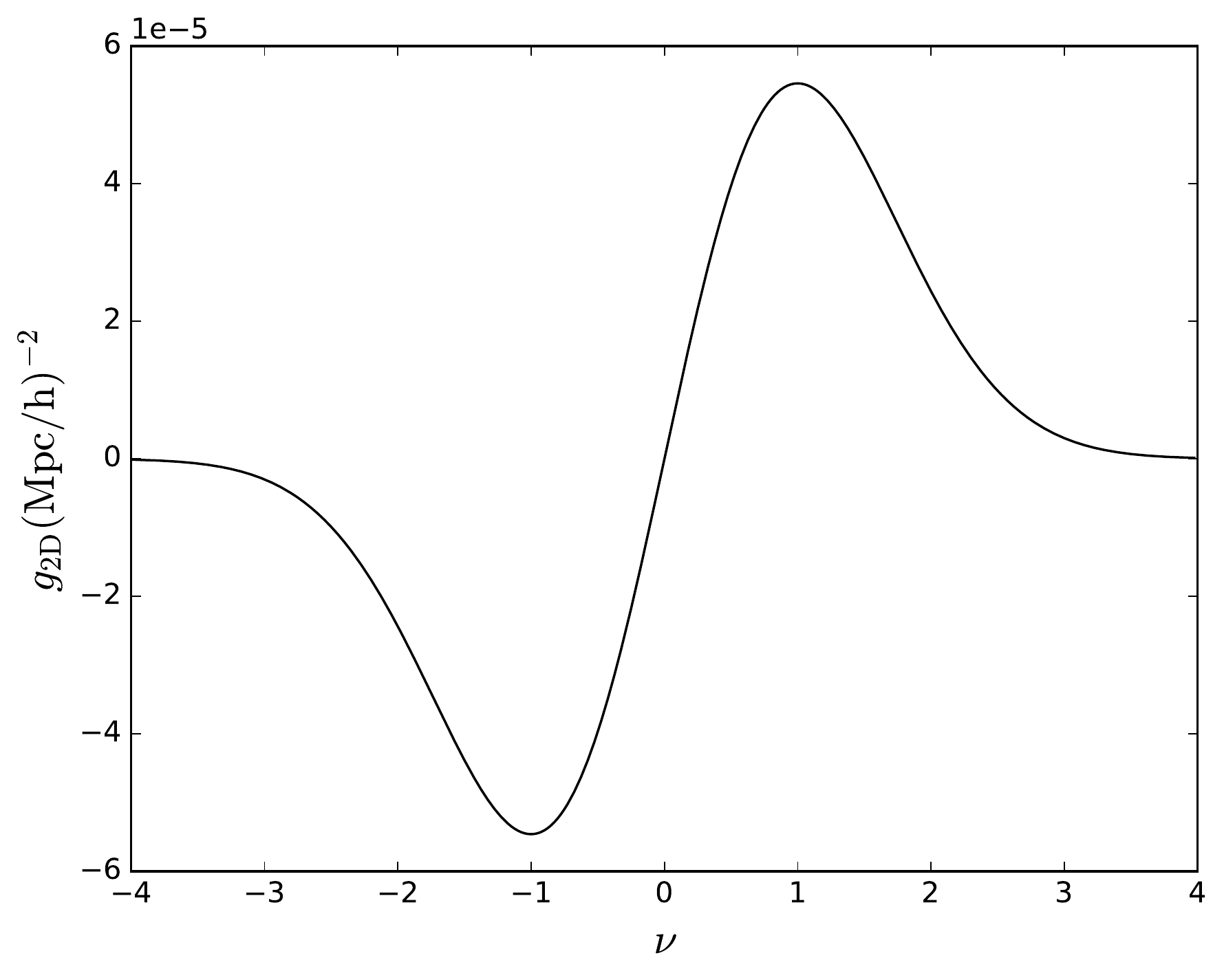}
  \includegraphics[width=0.45\textwidth]{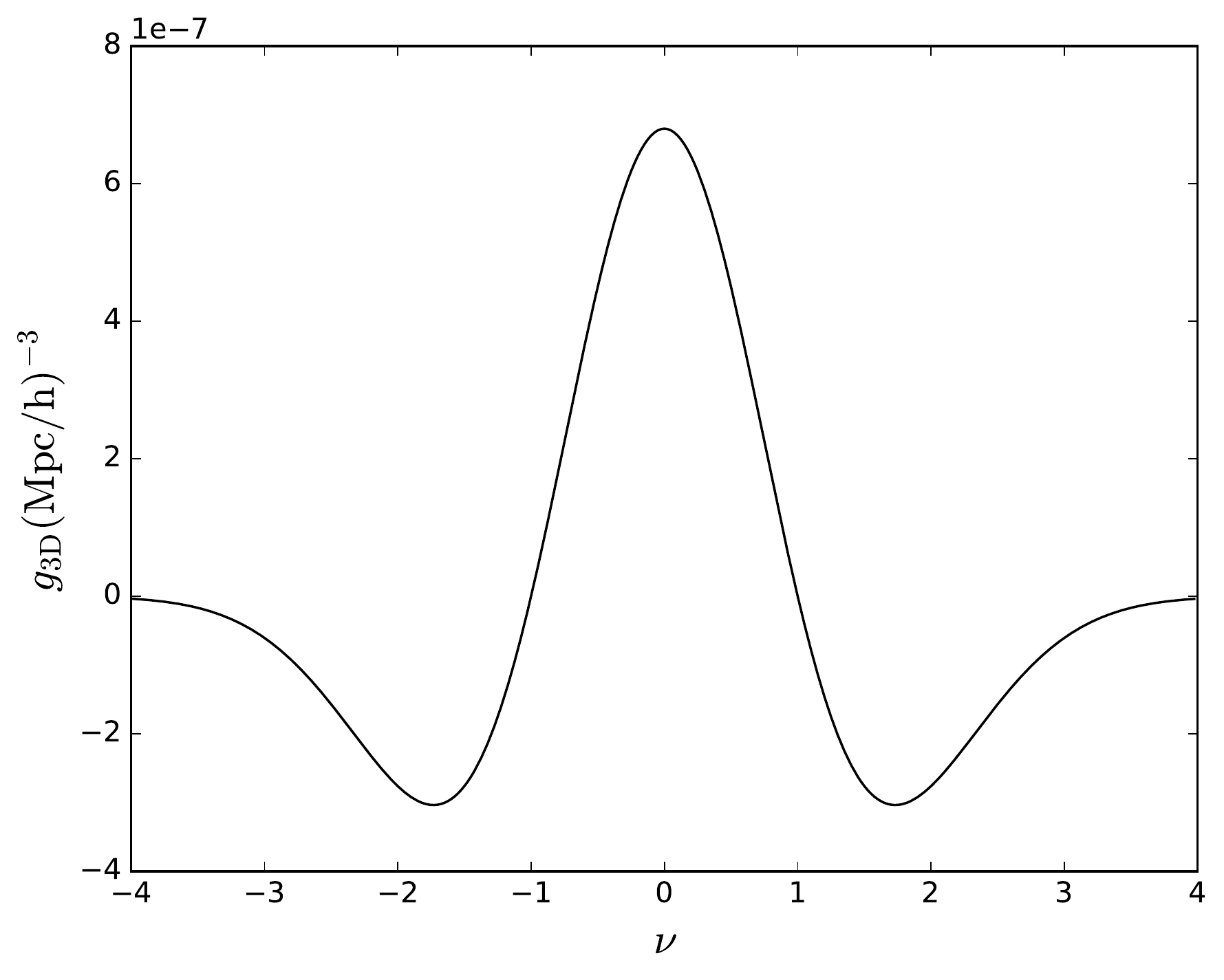}\\
  \caption{The two (top panel) and three (bottom panel) dimensional genus curves as a function of normalised density threshold $\nu$. We have assumed a Gaussian field with a $\Lambda$CDM power spectrum using Horizon Run 4 fiducial parameters. The two(three) dimensional genus is anti-symmetric(symmetric) with respect to $\nu \to -\nu$. The shape of the genus curve is fixed for a Gaussian field, and only the amplitude carries information. }
  \label{fig:gen_gauss}
\end{figure}

The genus per unit volume of the full three dimensional field as a function of excursion set threshold $\nu$ is given by \citep{1970Ap......6..320D,Adler,Gott:1986uz,Hamilton:1986}

\begin{eqnarray} \label{eq:gg} & &  g_{\rm 3D}(\nu) = {1 \over 8\pi^{2}} \left({\Sigma_{1}^{2} \over 3\Sigma_{0}^{2}}\right)^{3/2} \left(1 - \nu^{2} \right) e^{-\nu^{2}/2} , \\
\nonumber & & \Sigma_{0}^{2} = \langle \delta^{2}_{\rm 3D} \rangle  ,   \qquad  \Sigma_{1}^{2} = \langle |\nabla \delta_{\rm 3D} |^{2} \rangle  . \end{eqnarray} 

\noindent where $\Sigma_{0,1}$ are the cumulants of the three dimensional field. Similarly for the genus of two dimensional slices, we can write the genus per unit area as \citet{1989ApJ...345..618M}

\begin{eqnarray} \label{eq:gg2d} & &  g_{\rm 2D}(\nu) = {1 \over 2(2\pi)^{3/2}} {\sigma_{1}^{2} \over \sigma_{0}^{2}} \nu e^{-\nu^{2}/2} , \\
\nonumber & & \sigma_{0}^{2} = \langle \delta_{\rm 2D}^{2} \rangle  , \qquad  \sigma_{1}^{2} = \langle |\nabla \delta_{\rm 2D} |^{2} \rangle  . \end{eqnarray} 

\noindent where $\sigma_{0,1}$ are now cumulants of the two dimensional field. We can relate the two and three-dimensional cumulants to the power spectrum as

\begin{eqnarray}
\label{eq:s03} & & \Sigma_{0}^{2} = \int d^{3} k e^{-k^{2}\Lambda_{\rm G}^{2}/2} P_{\rm 3D}(k)   , \\
\label{eq:s13} & & \Sigma_{1}^{2} = \int d^{3} k e^{-k^{2}\Lambda_{\rm G}^{2}/2} k^{2} P_{\rm 3D}(k)   , \\
\label{eq:s02} & & \sigma_{0}^{2} = \int d^{2} k_{\perp} e^{-k_{\perp}^{2}R_{\rm G}^{2}/2} P_{\rm 2D}(k_{\perp})   , \\
\label{eq:s12} & & \sigma_{1}^{2} = \int d^{2} k_{\perp} k_{\perp}^{2} e^{-k_{\perp}^{2}R_{\rm G}^{2}/2} P_{\rm 2D}(k_{\perp}), 
  \end{eqnarray} 

\noindent where $\Lambda_{\rm G}$ is the three dimensional Gaussian smoothing scale and $R_{\rm G}$ the two dimensional Gaussian smoothing scale perpendicular to the line of sight. In what follows we focus on the two dimensional genus, but return to the three dimensional counterpart in section \ref{sec:amp}. We exhibit two and three dimensional genus curves $g_{\rm 2D}(\nu)$ and $g_{\rm 3D}(\nu)$ as a function of normalised overdensity threshold $\nu$ in Figure \ref{fig:gen_gauss}, where we take a fiducial $\Lambda$CDM power spectrum with parameters listed in table \ref{tab:ii}. The shape of the genus curves is fixed and both possess symmetry about $\nu = 0$.

\begin{table}
\begin{center}
 \begin{tabular}{||c | c ||}
 \hline
 Parameter & Fiducial Value \\ [0.5ex] 
 \hline\hline
 $\Omega_{\rm mat}$ & $0.26$   \\ 
 \hline
 $\Omega_{\Lambda}$ & $0.74$   \\ 
 \hline
 $n_{\rm s}$ & $0.96$   \\ 
 \hline
 $\sigma_{8}$ & $0.794$   \\
 \hline
 $\Delta$ & $60 {\rm Mpc}/h$   \\
 \hline
 $\Lambda_{\rm G}$ & $20 {\rm Mpc}/h$   \\
 \hline
 $R_{\rm G}$ & $15 {\rm Mpc}/h$   \\
 \hline
\end{tabular}\label{tab:ini}
\caption{\label{tab:ii}Fiducial parameters used in the Horizon Run 4 simulation, and the parameters used to calculate the genus in this work. $\Delta$ is the thickness of our two dimensional slices of the density field, and $R_{\rm G}, \Lambda_{\rm G}$ are the two and three dimensional smoothing scales used when applying a Gaussian smoothing kernel to the density fields.  }
\end{center} 
\end{table}

Let us consider measuring the genus of the dark matter field at different redshifts. We take $P_{\rm 3D}$ to be the standard linear $\Lambda$CDM power spectrum. If we assume linear evolution of the field from some high redshift initial condition, then the power spectrum retains its shape with redshift - only its amplitude changes through the linear growth factor. However, since the genus is a measure of ratios of cumulants, it follows that the genus amplitude should be constant when measured at different epochs, and insensitive to both the growth rate and also linear galaxy bias. This argument is also true in the mildly non-linear regime - in \citet{Matsubara:1994we,Matsubara:2000mw,Pogosyan:2009rg,Gay:2011wz,Codis:2013exa}, a series of papers modeled departures from Gaussianity as an expansion of the genus in the parameter $\sigma_{0}$. To linear order in $\sigma_{0}$ one can write

\begin{eqnarray} \nonumber & &  g_{\rm 2D}(\nu_{\rm A}) = A e^{-\nu_{\rm A}^{2}/2} \left[ H_{1}(\nu_{\rm A})+ \left[ {2 \over 3} \left( S^{(1)} - S^{(0)}\right) H_{2}(\nu_{\rm A}) + \right. \right. \\
\label{eq:mat1} & & \qquad \qquad \qquad + \left. \left. {1 \over 3} \left(S^{(2)} - S^{(0)}\right) \right] \sigma_{0} + {\cal O}(\sigma_{0}^{2}) \right] , \end{eqnarray} 

\noindent where $A$ is the amplitude of the genus and the skewness parameters $S^{(0)}, S^{(1)}, S^{(2)}$ are related to the three point statistics of the density field as

\begin{equation} \label{eq:sk1}  S^{(0)} = {\langle \delta^{3} \rangle \over \sigma_{0}^{4}} , \, S^{(1)} = - {3 \over 4} {\langle \delta^{2} (\nabla^{2} \delta) \rangle \over \sigma_{0}^{2} \sigma_{1}^{2}} ,\, 
 S^{(2)} = -3  {\langle (\nabla \delta . \nabla \delta)  (\nabla^{2} \delta) \rangle \over \sigma_{1}^{4}} . \end{equation} 

\noindent This expansion has been continued to arbitrary order in \citet{Pogosyan:2009rg}. $H_{i}(x)$ are Hermite polynomials, the first three of which are given by $H_{0}(x) = 1$, $H_{1}(x)= x$, $H_{2}(x)= x^{2}-1$. We have defined $\nu_{A}$ as the density threshold such that the excursion set has the same area fraction as a corresponding Gaussian field - 

\begin{equation}\label{eq:afrac} f_{A} = {1 \over \sqrt{2\pi}} \int^{\infty}_{\nu_{A}} e^{-t^{2}/2} dt , \end{equation}

\noindent where $f_{A}$ is the fractional area of the field above $\nu_{A}$. This choice of $\nu_{\rm A}$ parameterization eliminates the non-Gaussianity in the one-point function \citep{1987ApJ...319....1G,1987ApJ...321....2W,1988ApJ...328...50M}. The salient point that we can draw from equation ($\ref{eq:mat1}$) is that to linear order in $\sigma_{0}$ the genus amplitude, which is the coefficient of the $H_{1}$ Hermite polynomial, is unaffected by gravitational collapse. 

Returning to our Gaussian field example, let us take a three dimensional $\Lambda$CDM linear matter power spectrum $P_{\rm 3D}(k)$, generate a three dimensional field $\delta_{\rm 3D}$ and then take two dimensional slices $\delta_{\rm 2D}$ of thickness $\Delta = 60 {\rm Mpc}/h$. We smooth the two dimensional field using a Gaussian kernel with scale $R_{\rm G} = 15 {\rm Mpc}/h$  and calculate the genus per unit area for five different cosmological models as a function of redshift. From these curves (an example is shown in the top panel of Figure \ref{fig:gen_gauss}) we extract the amplitude - to do so  we measure the genus at $i_{\nu} = 50$ values of $\nu_{\rm A}$, equi-spaced between $-4  \le \nu_{\rm A} \le 4$. We assume that the genus curve can be expanded in terms of Hermite polynomials as follows

\begin{equation}\label{eq:fit}  \hat{g}_{\rm 2D}(\nu_{\rm A}) =   e^{-\nu_{\rm A}^{2}/2} \sum_{i} a_{i} H_{\rm i}(\nu_{\rm A}) , \end{equation} 

\noindent with constant $a_{i}$ coefficients. Then, we can use the orthogonality property of the Hermite polynomials 

\begin{equation} \label{eq:or1} \int_{-\infty}^{\infty} H_{\rm m}(x) H_{\rm n}(x) e^{-x^{2}/2} dx = \sqrt{2\pi}n! \delta_{\rm n m} \end{equation}

\noindent to extract the amplitude from the genus. We numerically integrate the measured genus curve over the range $-4 \le \nu_{\rm A} \le 4$ with window function $H_{\rm 1} = \nu_{\rm A}$, the output of which corresponds to the amplitude $A$, multiplied by a factor of $\sqrt{2\pi}$.

\begin{figure}
  \includegraphics[width=0.45\textwidth]{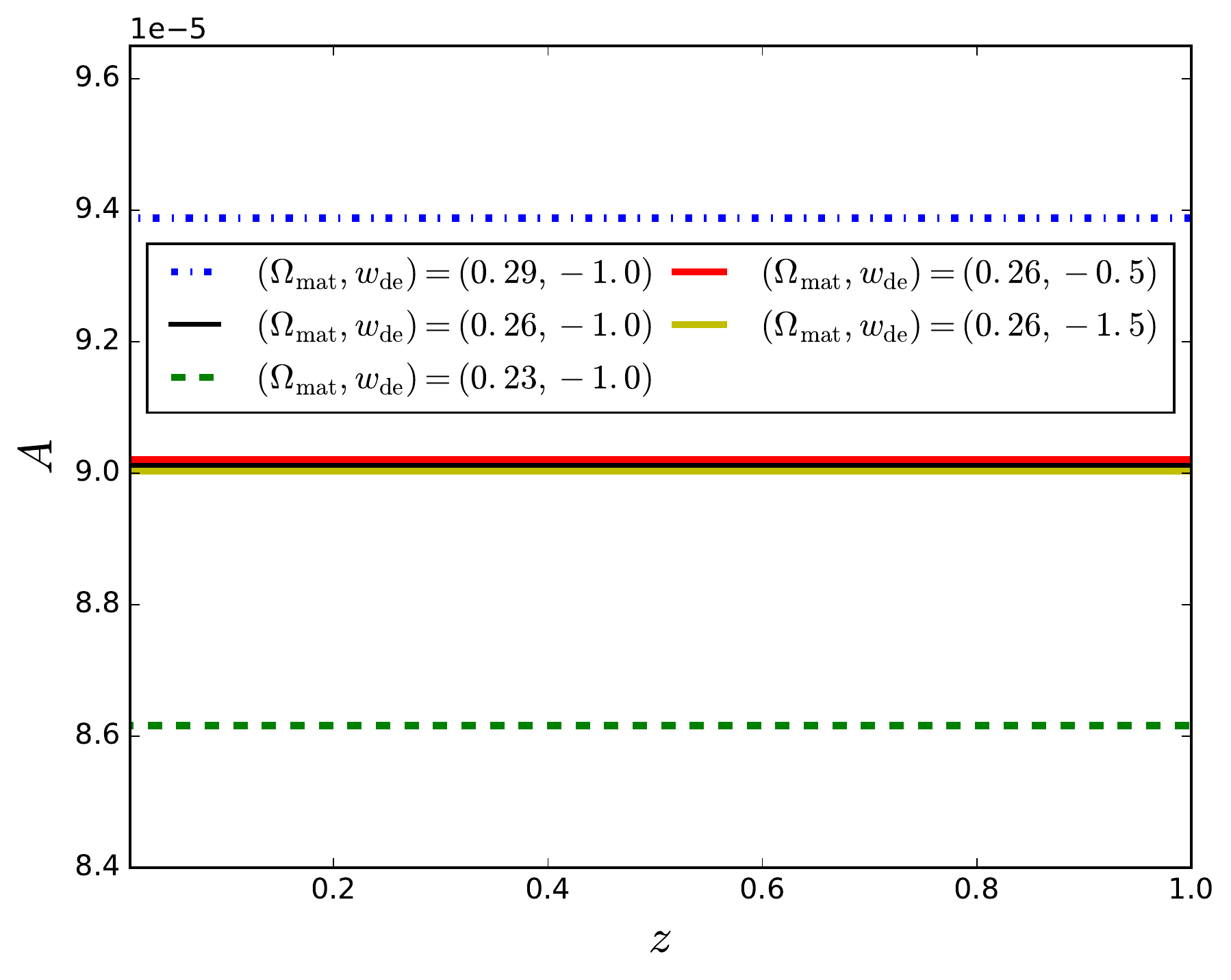}
  \includegraphics[width=0.45\textwidth]{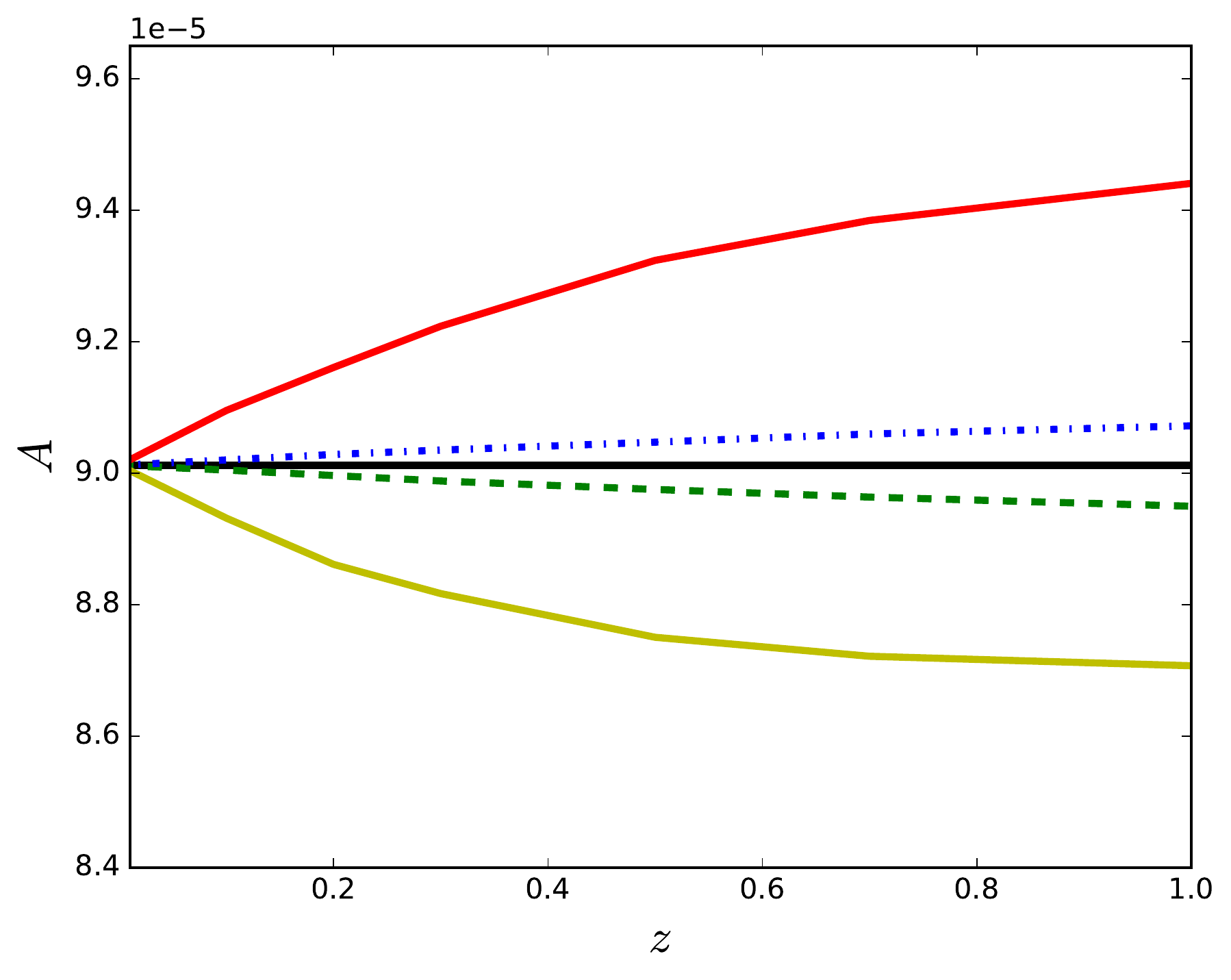}\\
  \caption{(Top panel) Predicted two dimensional genus amplitude (per unit area) for five different cosmological models. The genus amplitude is sensitive to $\Omega_{\rm mat}$ as this parameter modifies the shape of the power spectrum. However, if we assume that dark energy does not strongly cluster then its effect on the genus amplitude is negligible. (Bottom panel) If data is drawn from some `fiducial' underlying cosmology (here $(\Omega^{\rm (fid)}_{\rm mat},w^{\rm (fid)}_{\rm de})=(0.26,-1.0)$) but we measure the genus amplitude assuming a different cosmology, then the genus amplitude will evolve with redshift. We observe this effect for four different cosmological models. The equation of state of dark energy is principally significant when comparing low $z<0.1$ to high $z > 0.5$ redshift genus amplitudes - we exhibit strong evolution when we select $w_{\rm de}  = -0.5, -1.5$ (red and yellow lines respectively). }
  \label{fig:2}
\end{figure}

The results are shown in the top panel of Figure \ref{fig:2}. We calculate $A$ for five cosmological models $(\Omega_{\rm mat},w_{\rm de}) = (0.26,-1), (0.23,-1), (0.29,-1), (0.26,-0.5), (0.26,-1.5)$. The first of these is our `fiducial cosmology' in the sense that it is the one used to construct the Horizon Run dark matter simulations that we will introduce in section \ref{sec:sim}. We fix all other parameters to WMAP5 best fit values \citep{2009ApJS..180..306D}, and use the publicly available software CAMB\footnote{http://camb.info} \citep{Lewis:1999bs} to generate the matter power spectra $P_{\rm 3D}$ for these cosmologies, from which we calculate the amplitude of the genus over the redshift range $0 < z < 1$. One can observe that the genus amplitude is constant with $z$, and varies with $\Omega_{\rm mat}$ as this parameter modifies the location of the peak of the power spectrum. The genus amplitude is effectively independent of the dark energy equation of state  $w_{\rm de}$. Dark energy perturbations, whilst present for $w_{\rm de} \ne -1$, provide a subdominant effect on the matter power spectrum.

Next, we generate a density field from a power spectrum using some `fiducial cosmology' $(\Omega^{\rm (fid)}_{\rm mat},w^{\rm (fid)}_{\rm de})=(0.26,-1)$. Suppose that an observer is presented with this density field and is not privy to the cosmological model from which it is drawn. Then the observer will measure the amplitude of the genus at various redshifts assuming a particular set of cosmological parameters $(\Omega_{\rm mat}, w_{\rm de})$ to infer the distance redshift relation. Choosing an incorrect cosmology $(\Omega_{\rm mat}, w_{\rm de}) \ne (\Omega^{\rm (fid)}_{\rm mat},w^{\rm (fid)}_{\rm de})$ will effect our measurement of the genus amplitude in two ways - the smoothing scale $R_{\rm G}$ and the total area of the two dimensional slice will systematically evolve\footnote{Additional cosmological parameter dependence will enter via our choice of constant comoving shell thickness $\Delta$, although this effect is small when taking thick shells}. For a fixed volume snapshot density field that we are considering in this section, the smoothing scale $R_{\rm G}$ will be modified by a factor $\lambda = d_{\rm cm, X}(z)/d_{\rm cm, Y}(z)$, where $d_{\rm cm}$ is the comoving distance at redshift $z$ of the two dimensional slice and $X, Y$ are shorthand notation for the correct and incorrect cosmological models. In addition the genus amplitude per unit area will be modified by a factor of $\lambda^{2}$. These two effects do not precisely cancel as the field is not scale invariant. The result is that the genus amplitude will systematically evolve if we measure the genus using an incorrect cosmology when calculating distances. This evolution is exhibited in the bottom panel of Figure \ref{fig:2}. The amplitude now varies as a function of the equation of state of dark energy (red and yellow curves), due to the dependence of the smoothing scale $R_{\rm G}$ on the distance-redshift relation. These effects were first noted for the three dimensional genus in \citet{Park:2009ja}.

There are two ways that one could use the genus amplitude to generate cosmological parameter constraints. The first is to note that $A$ evolves with redshift if we select an incorrect cosmological model, and is constant only if we choose the correct set of parameters to infer the distance redshift relation. It follows that the genus amplitude can be used as a standard population, in the sense that we can make multiple measurements at different redshifts, and find the cosmology that minimizes the evolution. In our example (the bottom panel of Figure \ref{fig:2}) this is clearly the fiducial cosmology $(\Omega^{\rm (fid)}_{\rm mat},w^{\rm (fid)}_{\rm de})=(0.26,-1)$.

Alternatively, we can also directly use the magnitude of the amplitude. In this case, we would measure the genus of a density field using particular cosmological parameters $(\Omega_{\rm mat}, w_{\rm de})$ at different redshifts, effectively obtaining a curve such as in the bottom panel of Figure \ref{fig:2}. We then compare these amplitude measurements to the Gaussian expectation value corresponding to the same parameter set (the top panel of Figure \ref{fig:2}). The underlying cosmology is one which minimizes the difference between these two curves - in the figure that is again $(\Omega_{\rm mat},w_{\rm de})=(0.26,-1)$. 

If the curves in the bottom panel were constant in redshift, then we would find perfect degeneracy between $\Omega_{\rm mat}$ and $w_{\rm de}$ when comparing the measured genus amplitude values to their Gaussian expectation value. The (slight) redshift dependence breaks the degeneracy and allows us to potentially place constraints on the equation of state of dark energy. To maximize the constraining power of the method, we require a measurement of the genus at low redshift $z \sim 0.1$ to compare to the higher redshift value. This presents the dominant challenge of the method - we require an accurate measurement of the genus at low redshift, using a sufficiently large smoothing scale $R_{\rm G}$ such that we can assume that the expansion ($\ref{eq:mat1}$) of $g_{\rm 2D}$ in $\sigma_{0}$ is valid, and the non-linear correction to the amplitude is subdominant. 

Schematically, these two methods involve minimizing the following $\chi^{2}$ distributions 

\begin{eqnarray} \label{eq:chi1} & &  \chi_{\rm evo}^{2} = \sum_{i} {\left( A_{i}(z_{i},\Omega_{\rm mat},w_{\rm de}) - A_{0} \right)^{2} \over \sigma_{i}^{2} } \\ 
\label{eq:ch} & &  \chi_{\rm mag}^{2} = \sum_{i} {\left( A_{i}(z_{i},\Omega_{\rm mat},w_{\rm de}) - A_{\rm G}(\Omega_{\rm mat}) \right)^{2} \over \sigma_{i}^{2}} \end{eqnarray}

\noindent where the $i$ subscript runs over the redshift slices at which we measure the genus, $A_{i}(z_{i},\Omega_{\rm mat},w_{\rm de})$ are the measured amplitudes using assumed cosmological parameters $\Omega_{\rm mat},w_{\rm de}$ and $\sigma_{i}$ are the rms statistical fluctuations of the measurements, estimated using mock density fields. If we consider solely the evolution of the genus, then we should minimize $\chi_{\rm evo}^{2}$, where $A_{0}$ is an unimportant constant that we fit. If we wish to additionally use the magnitude of the amplitude, then we minimize $\chi^{2}_{\rm mag}$, where $A_{\rm G}$ is the Gaussian amplitude given parameters $(\Omega_{\rm mat}, w_{\rm de})$. As we will show, before we use either method we must first carefully account for systematics that might otherwise bias the constraints.

The quantity $A_{\rm G}(\Omega_{\rm mat})$ can be estimated either by calculating the cumulants $\sigma_{0,1}$ by direct integration of the linear matter power spectrum according to equations ($\ref{eq:s02},\ref{eq:s12}$), or by generating Gaussian random fields and measuring either $\sigma_{0,1}$ or $A_{\rm G}$ directly. We adopt the latter approach, and generate discrete, three dimensional Gaussian random fields in Fourier space drawn from a linear, $\Lambda$CDM dark matter power spectrum $P_{\rm 3D}(k)$. $P_{\rm 3D}(k)$ is calculated for a particular cosmological parameter set using CAMB. The resulting three dimensional field $\delta_{ijk}^{\rm G}$ is chosen to occupy a $V = (3150 {\rm Mpc}/h)^{3}$ box and discretized using resolution $\delta x = \delta y = \delta z = 3.1 {\rm Mpc}/h$. The field is sliced into two-dimensional subsets of width $\Delta = 60 {\rm Mpc}/h$ and smoothed in the plane using a Gaussian kernel with scale $R_{\rm G} = 15 {\rm Mpc}/h$. Genus curves are generated from the two dimensional density field using standard methods \citep{Appleby:2017ahh}. From $g_{\rm 2D}(\nu_{\rm A})$, amplitudes are calculated by integrating $\nu_{\rm A} g_{\rm 2D}(\nu_{\rm A})$ over the threshold range $-4 < \nu_{\rm A} <4$, using the orthogonality condition ($\ref{eq:or1}$) of the Hermite polynomials. The average amplitude of $N_{\rm slice} = 50$ slices is taken to be $A_{\rm G}(\Omega_{\rm mat})$. This procedure is repeated for $\Omega_{\rm mat}$ over the range $0.1 < \Omega_{\rm mat} < 0.8$ with resolution $\Delta \Omega_{\rm mat} = 10^{-3}$ - the result is an array of $A_{\rm G}(\Omega_{\rm mat})$ values. $A_{\rm G}(\Omega_{\rm mat})$ is then reconstructed for any $\Omega_{\rm mat}$ value by spline interpolation of the $(\Omega_{\rm mat}, A_{\rm G})$ curve. The resulting $A_{\rm G}(\Omega_{\rm mat})$ is the Gaussian expectation value for a dark matter field with no shot noise contribution. Note that to construct $A_{\rm G}$, we have neglected dark energy perturbations and assumed that any dependence on $w_{\rm de}$ is negligible.

\section{Can we use the genus amplitude to constrain the dark energy equation of state?}
\label{sec:genevo}

We first address the question of whether we can use purely the redshift dependence of the genus amplitude (cf the bottom panel of Figure \ref{fig:2}) for cosmological parameter estimation. This test has the desirable property that the resulting parameter constraints would be independent of any model assumptions regarding the growth of structure, and would not depend on the genus amplitude (only the constancy of its value). 

For illustrative purposes we apply our method to a Gaussian field. We measure the genus at four different redshifts - $z=(0.1, 0.2, 0.5, 1)$. We generate large scale Gaussian random fields in snapshot boxes of comoving volume $3150^{3} \left({\rm Mpc}/h\right)^{3}$ using a fiducial cosmological model $(\Omega^{\rm (fid)}_{\rm mat},w^{\rm (fid)}_{\rm de}) = (0.26, -1)$. We then take $N_{\rm slice} = 50$ slices of the field of thickness $\Delta = 60 {\rm Mpc}/h$. We set a fiducial smoothing scale perpendicular to the line of sight $R_{\rm G} = 15 {\rm Mpc}/h$.

We take three different cosmological models - $(\Omega_{\rm mat}, w_{\rm de}) = (0.26, -0.5), (0.26, -1), (0.26, -1.5)$, and measure the genus per unit area $g_{\rm 2D}$ at each redshift assuming a distance redshift relation $d_{\rm cm}(z_{\rm i},\Omega_{\rm mat},w_{\rm de})$. As discussed in the previous section, the choice of cosmological parameters will modify the analysis in two ways - we smooth using modified scale $R_{\rm G} \lambda$ and adjust the genus per unit area $g_{\rm 2D}$ by a factor of $\lambda^{2}$.

We calculate the genus per unit area  of the $N_{\rm slice} = 50$ density field slices at each redshift, and use the genus curves $g_{\rm 2D}(\nu)$ to infer the amplitude. The mean and rms fluctuations of the amplitude measurements at each redshift are denoted $\bar{A}_{i}$ and $\hat{\sigma}_{i}$ respectively, where $i$ subscripts run over the four redshifts. We then weight $\hat{\sigma}_{i}$ by a factor

\begin{equation}\label{eq:si} \sigma_{i} =  \sqrt{ 3150^{2} \over 4\pi d^{2}_{\rm cm}(z_{i},\Omega^{\rm (fid)}_{\rm mat},w^{\rm (fid)}_{\rm de})} \hat{\sigma}_{i} \end{equation}

\noindent which simply scales our error bar by the effective area available to us at each redshift. We use the fiducial cosmology for this weighting. A factor of the observed sky fraction $f_{\rm sky}$ can be included in the denominator of $(\ref{eq:si}$), but we fix $f_{\rm sky} = 1$. 

\begin{figure}
  \includegraphics[width=0.45\textwidth]{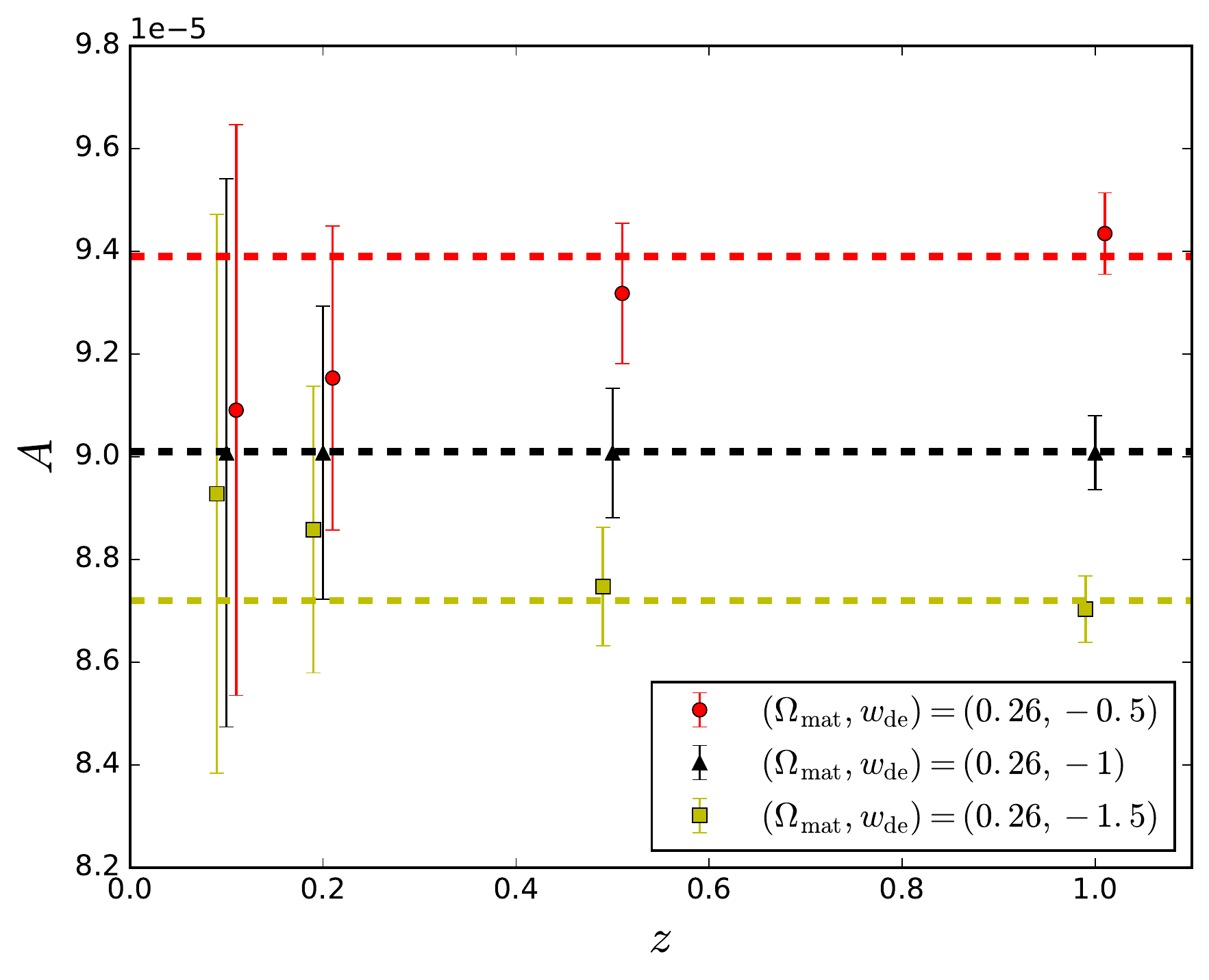}
  \caption{Genus amplitudes of a density field generated from the fiducial cosmological model $(\Omega^{\rm (fid)}_{\rm mat}, w^{\rm (fid)}_{\rm de})=(0.26,-1)$ then measured by assuming three different cosmological models $(\Omega_{\rm mat}, w_{\rm de})=(0.26,-0.5), (0.26,-1), (0.26,-1.5)$ (red points, black triangles and yellow squares). The error bars are estimated statistical uncertainty given a smoothing scale $R_{\rm G} = 15 {\rm Mpc}/h$ and all sky data at the redshifts. We note the large error bar at low $z$. }
  \label{fig:chi1}
\end{figure}

We exhibit $\bar{A}_{i}$ and $\sigma_{i}$ error bars as a function of $z$ in Figure \ref{fig:chi1}. We observe the correct behaviour - an increasing/decreasing amplitude for $w_{\rm de} = (-0.5, -1.5)$ respectively. However, the error bar on the low redshift measurement is larger than the signal.

This analysis highlights the principle difficulty associated with the measurement - the large statistical fluctuations at low redshift. Therefore, it is very important to increase the statistics at low redshifts by increasing the sample volume of the nearby Universe if the evolution of the genus is to be used to constrain cosmology. In the following section we consider how our ability to constrain $w_{\rm de}$ improves if we use tomography to reduce the statistical fluctuations associated with our measurement.

\section{Application to Simulated data - Horizon Run 4}\label{sec:sim}

To perform a tomographic analysis we transition to a more realistic scenario. First, we introduce a density field that better represents the low redshift Universe, which is both non-Gaussian and observed via biased tracers. Throughout this work we use the Horizon Run 4 simulation \citep{Kim:2015yma}. It is a dense, cosmological scale $N$-body simulation that gravitationally evolved $N=6300^{3}$ particles in a volume of $V=(3150 \, {\rm Mpc}/h)^{3}$. The simulation uses a modified version of GOTPM code and initial conditions are calculated using second order Lagrangian perturbation theory. The cosmological parameters used are given in table \ref{tab:ii}, and details of the simulation can be found in \citet{Kim:2015yma}. Both dark matter particle and mock galaxy data is available, in the form of snapshot data and a single all-sky lightcone to $z = 1.4$. Details of the numerical implementation, and the method by which mock galaxy catalogs are constructed can be found in \citet{Hong:2016hsd}. Mock galaxies are defined using the most bound halo particle-galaxy correspondence scheme. Survival time of satellite galaxies after merger is calculated by adopting the merger timescale model described in \citet{Jiang:2007xd}. 

We use mock galaxy lightcone as our data sample. We consider the redshift range $0.1 < z < 1$, generating $N_{\rm shell} = 36$ concentric redshift shells of thickness $\Delta = 60 {\rm Mpc}/h$. The center of the $i^{\rm th}$ redshift shell is denoted $\hat{z}_{i}$ for $1 \le i \le N_{\rm shell}$. We fix the comoving number density of mock galaxies as $\bar{n} = 10^{-3} \left( {\rm Mpc}/h \right)^{-3}$, by adopting a minimum galaxy mass cut in each of our tomographic bins. Throughout this work we use fiducial scale $R_{\rm G} = 15 {\rm Mpc}/h$ to smooth perpendicular to the line of sight and adopt the $\nu_{\rm A}$ parameterization of the excursion sets to minimize one-point, non-Gaussian contributions to the genus curve \citep{1987ApJ...319....1G,1987ApJ...321....2W,1988ApJ...328...50M}. We bin galaxies according to their angular positions using an equal area HEALPix\footnote{http://healpix.sourceforge.net} pixelation of the unit sphere \citep{Gorski:2004by}. We denote the number of galaxies at each grid point on the sphere as $n_{j}$, where $j$ runs over the total number of pixels. Defining $\bar{n}_{\rm p}$ as the average number of galaxies in a pixel, we define the density field $\delta_{j} = (n_{j} - \bar{n}_{\rm p})/\bar{n}_{\rm p}$. 

We smooth this two dimensional map with angular scale $\theta_{{\rm G}, i} = R_{\rm G} / d_{\rm cm}(\hat{z}_{i},\Omega_{\rm mat}, w_{\rm de})$, where $d_{\rm cm}$ is the comoving distance to the $i^{\rm th}$ shell. We adopt $N_{\rm pix} = 12 \times N_{\rm side}^{2}$ pixels, with $N_{\rm side} =1024$, and calculate the genus of this field. This pixel number was chosen to ensure that pixel effects remain negligible for all $z \le 1$. When calculating the genus on a sphere, there are two approaches that one can take. One is to project the density field from the shell onto the plane, using a conformal mapping. Any angle preserving projection will conserve the genus. Alternatively, one can calculate the genus directly on the sphere using the method constructed in \citet{Schmalzing:1997uc}. We adopt the latter approach - an algorithm was developed in \citet{Schmalzing:1997uc} and adopted by the authors in \citet{Appleby:2017ahh}, and we direct the reader to these works for details. The genus per comoving area is defined as $g_{\rm 2D} = G_{\rm 2D}/d^{2}_{\rm cm}(\hat{z}_{i},\Omega_{\rm mat}, w_{\rm de})$, where $G_{\rm 2D}$ is the genus per unit area on the unit two-sphere. 

Our choice of cosmological model $(\Omega_{\rm mat},w_{\rm de})$ enters into the calculation in two ways - when converting constant physical scale $R_{\rm G}$ to a corresponding angular scale $\theta_{\rm G}$ and when calculating the genus per unit area $g_{\rm 2D}$ from the genus on the unit sphere $G_{\rm 2D}$. Cosmology will also enter when we consider the effect of redshift space distortions - this will be discussed in section \ref{sec:rsd}. For now we study the density field in real space. Note that we no longer require the conversion factor $\lambda$ between different cosmologies, as we are not calculating the genus for a fixed box size - at no point in our analysis do we assume a `fiducial' cosmology. All distances are calculated using the cosmology adopted.

After measuring the genus curves and numerically integrating over the $\nu_{\rm A}$ range, the result is a set of $N_{\rm shell}=36$ amplitude measurements, one at each redshift - $A_{i}$. To reduce the statistical error we take every $N_{\rm a}=5$ adjacent redshift bin measurements, and calculate the mean genus amplitude. The standard error of this measurement is then $\bar{\sigma}_{j} = \sigma_{j}/\sqrt{N_{\rm a}}$, where $\sigma_{j}$ is the standard deviation of the five measurements. The result is seven effectively independent values of the genus amplitude $\bar{A}_{j}$, which are themselves a mean measurement of five adjacent shells. Here we have neglected the correlation between shells - in reality they will be correlated by both large scale Fourier modes and contamination due to photometric redshift errors. 

\begin{figure}
  \includegraphics[width=0.48\textwidth]{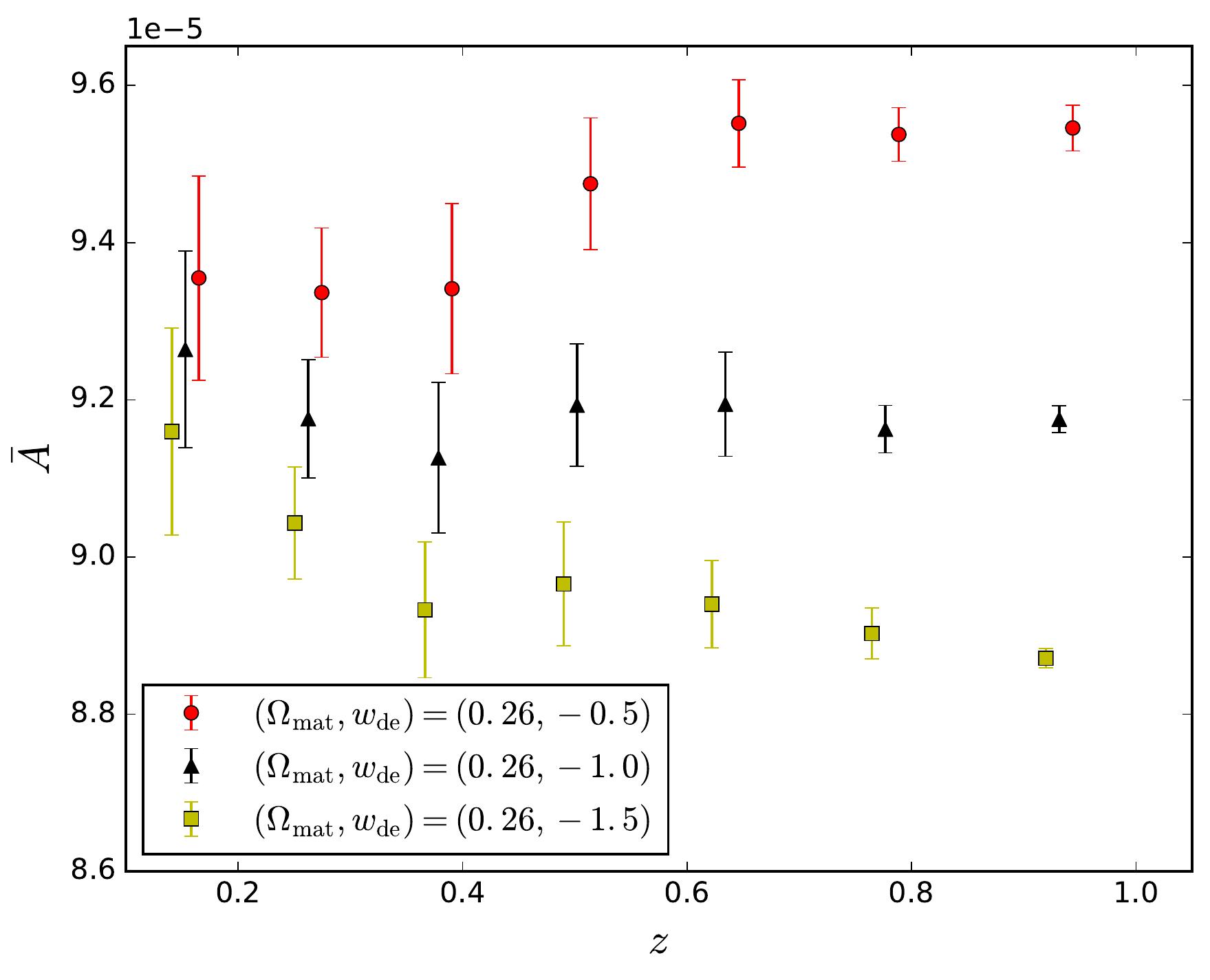}\\
  \caption{The mean tomographic measurements of the genus amplitude $\bar{A}_{j}$ obtained from the Horizon Run 4 lightcone data, as described in the text, using three different cosmological models to infer the distance redshift relation - $(\Omega_{\rm mat}, w_{\rm de})=(0.26,-0.5), (0.26,-1), (0.26,-1.5)$ (red points, black triangles and yellow squares). We observe similar behaviour to the Gaussian field as presented in Figure \ref{fig:chi1} - evolution when the incorrect cosmological model is used. The statistical significance of the signal has increased as a result using the average amplitude over five successive bins.  }
  \label{fig:lc}
\end{figure}

In Figure \ref{fig:lc} we exhibit the mean amplitude $\bar{A}_{j}$ and $\bar{\sigma}_{j}$ as a function of redshift for our mock galaxy lightcone data. As in Figure \ref{fig:chi1}, we calculate the genus amplitude assuming three different cosmological models - $(\Omega_{\rm mat}, w_{\rm de})=(0.26,-0.5), (0.26,-1), (0.26,-1.5)$ (red points, black triangles and yellow squares). The data points are located at the mean redshift $\bar{z}_{j}$ of the five adjacent redshift shells over we calculate the average $\bar{A}_{j}$.  

We find similar behaviour to the Gaussian case - evolution in $\bar{A}$ when we choose the incorrect cosmology to infer distances. The data points are consistent with no evolution with redshift over the range probed when using the correct underlying cosmology in the distance redshift relation. Calculating the average amplitude over multiple shells has reduced the statistical error on the measurements, and has produced a statistically significant signal.

Also note that the genus amplitude from the lightcone data (exhibited in Figure \ref{fig:lc}) is systematically higher than the Gaussian expectation value for the same cosmology (cf Figure \ref{fig:chi1}). This is principally due to shot noise, and will be relevant when we try to extract information from the magnitude of the genus amplitude. If we only use the redshift evolution then the absolute value of $\bar{A}$ is irrelevant and shot noise will not bias our statistics, subject to the condition that we fix a constant number density in each shell. 

The condition that sampling noise does not vary appreciably with redshift is crucial in our analysis. Galaxy surveys often produce flux-limited samples, so when we apply the method to actual data we must adopt an absolute magnitude cut in each redshift slice. This allows us to resolve the same large scale structures in each shell with roughly equal signal to noise. There are many examples in the literature where such volume-limited samples are constructed out of raw observed samples. Recent examples can be found in \cite{Choi:2010sx,Kim:2011ab,Choi:2013eej,Hwang:2016yme}. A luminosity evolution correction is applied to make constant number density samples in \cite{Choi:2010sx,Choi:2013eej}, which is equivalent to applying a redshift-dependent luminosity cut.

We have shown that the signal discussed in section \ref{sec:genevo} can be extracted from a galaxy catalog, and furthermore can be amplified by combining information from successive redshift shells to reduce the statistical fluctuations. However, thus far we have considered the density field in real space. In actuality the field will be observed in redshift space and Redshift Space Distortion (RSD) can affect both the magnitude and evolution of the genus. Before considering parameter constraints that can be generated, we consider the two dominant sources of systematic error that we must account for.

\section{Systematic Effects} 
\label{sec:sys}

We now consider the two main sources of contamination associated with the signal observed in Figure \ref{fig:lc} - redshift space distortion and small scale effects such as shot noise. We note that RSD is not strictly a systematic, and in fact by modeling its effect on the genus we can actually place additional constraints on the growth rate. We return to this point in section \ref{sec:amp}.

\subsection{Redshift Space Distortion} 
\label{sec:rsd}

In redshift space, the peculiar velocity along the line of sight affects the genus in both the linear and non-linear regimes. One of the advantages of taking thick slices $\Delta = 60 {\rm Mpc}/h$ of the three dimensional field is that non-linear Finger of God effects should be mitigated. However, the linear Kaiser effect will affect the genus amplitude. 

To leading order in an expansion of the density field, the effect of redshift space distortion on the two dimensional genus is given by a fractional amplitude shift $a^{\rm (2D)}_{\rm RSD}$ as \citep{1996ApJ...457...13M, Matsubara:2000mw, Codis:2013exa, Appleby:2017ahh}

\begin{eqnarray} \label{eq:gen_rsd} & & g_{\rm 2D}^{\rm RSD}(\nu,\theta_{\rm s}) = a^{\rm (2D)}_{\rm RSD}  g_{\rm 2D}^{\rm real}(\nu) , \end{eqnarray} 

\noindent where 

\begin{equation}\label{eq:amp_rsd} a^{\rm (2D)}_{\rm RSD} = {3 \over 2} \sqrt{\left( 1 - {C_{1} \over C_{0}}\right)\left[ 1 - {C_{1} \over C_{0}} + \left({3C_{1} \over C_{0}} -1 \right) \cos^{2}(\theta_{\rm s})\right]}  , \end{equation}

\noindent and 

\begin{equation} {C_{1} \over C_{0}} = {1 \over 3} {1 + 6 \beta /5 + 3 \beta^{2}/7 \over 1 + 2 \beta/3 +  \beta^{2}/5} . \end{equation}

\noindent Here, $\theta_{\rm s}$ is the angle between the plane and the line of sight (in this work we fix $\theta_{\rm s} = \pi/2$) and $\beta = f/b$ is the redshift space distortion parameter. $b$ is the (possibly redshift dependent) tracer bias, $f = \dot{D}/(HD)$ is the growth factor, and $D$ is the linear growth rate. For a $w$CDM model, we can write $f \simeq \Omega^{\gamma}_{\rm m}(z)$, where $\gamma = 3(1-w_{\rm de})/(5-6 w_{\rm de})$ and $\Omega_{\rm m}(z)=\Omega_{\rm mat}(1+z)^{3}H_{0}^{2}/3H^{2}$. The genus amplitude now acquires additional dependence on the cosmological model via both $f$ and the linear bias $b$. The bias dependence is particularly troubling although it is inevitable that statistics in redshift space will depend on the galaxy population under consideration and their place in the cosmic web, as the velocity of the tracers depend on the local gravitational potential.

For fixed $\Omega_{\rm mat}$ and $b$, $a^{\rm (2D)}_{\rm RSD}$ generically decreases with redshift. However, this picture is complicated by the fact that any sample of tracers that we use will have a bias that evolves with redshift; $b = b(z)$. Therefore $a^{\rm (2D)}_{\rm RSD}$ can increase, decrease or remain constant with redshift for any given cosmology.

\begin{figure}
  \includegraphics[width=0.45\textwidth]{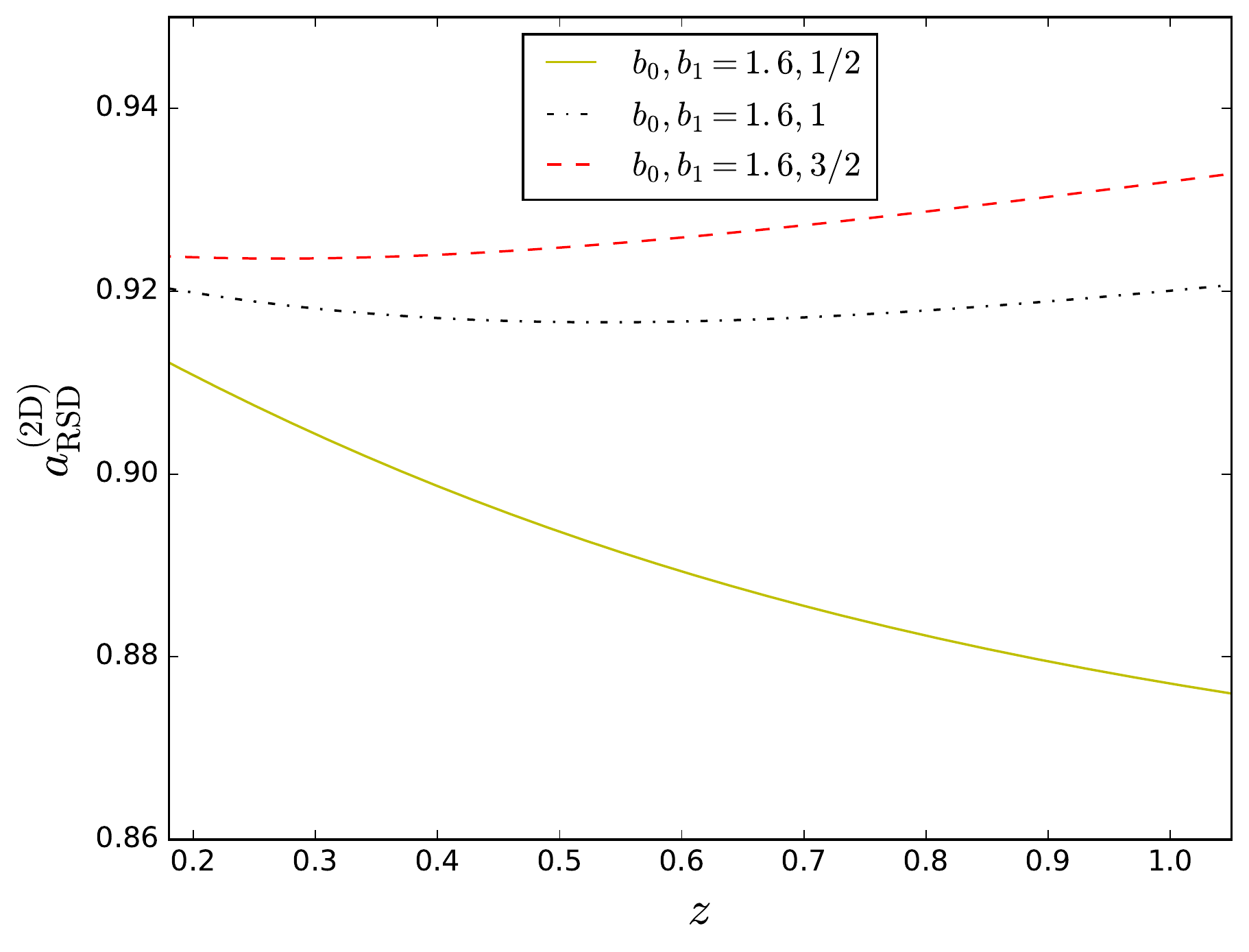}\\
  \includegraphics[width=0.45\textwidth]{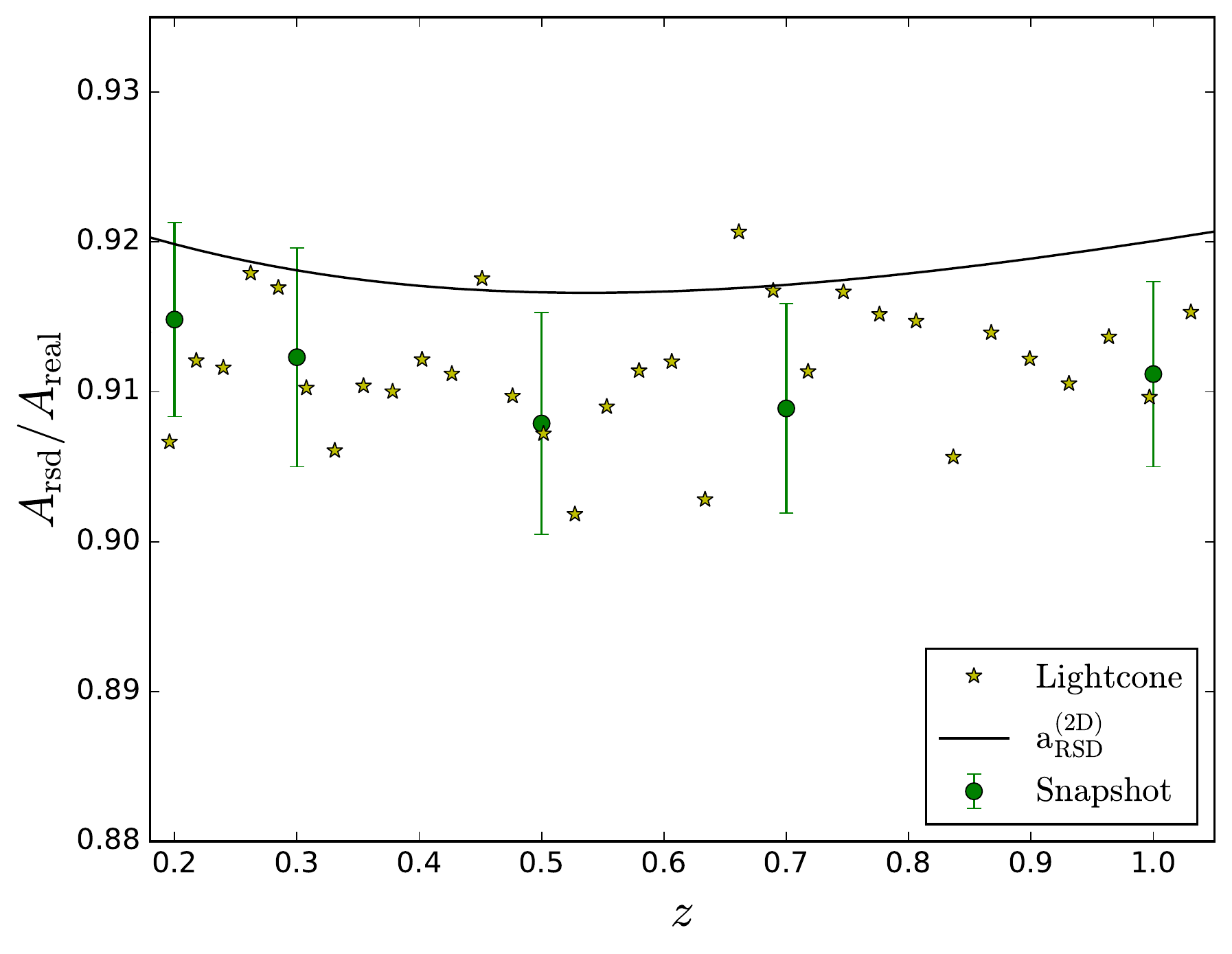}\\
  \caption{(Top panel) The function $a^{\rm (2D)}_{\rm RSD}$ defined in the text, as a function of redshift. We have fixed $\Omega_{\rm mat} = 0.26$ and introduced a redshift dependent bias $b(z) = b_{0} + b_{1}z$. We exhibit the function for various values of $b_{1}$, fixing $b_{0}=1.6$. The data set that we use is fit by $b(z) = 1.6 + z$, corresponding to the black dot-dashed line. $a^{\rm (2D)}_{\rm RSD}$ can increase, decrease or stay approximately constant over the redshift range probed, subject to the behaviour of $b(z)$. (Bottom panel) For our mock galaxy sample we can calculate the genus amplitude of the field in real $A_{\rm real}$ and redshift space $A_{\rm rsd}$. The green dots/error bars are the mean and rms fluctuations of the ratio $A_{\rm rsd}/A_{\rm real}$ for snapshot data at $z=0.2,0.3,0.5,0.7,1.0$. The yellow stars are the same quantity for our lightcone shells. We observe agreement between the two data sets. The solid black line is the linear theory theoretical prediction for $a_{\rm rsd}$ assuming bias $b(z) = 1.6 + z$. }
  \label{fig:rsd2}
\end{figure}

We fix $\Omega_{\rm mat} = 0.26$, adopt a bias model of $b = b_{0} + b_{1} z$ and show $a^{\rm (2D)}_{\rm RSD}$ as a function of $z$ for various $b_{1}$ values in the top panel of Figure \ref{fig:rsd2}, fixing $b_{0} = 1.6$. We choose this value of $b_{0}$ as our mock galaxy sample has a bias that is well fit by $b(z) = 1.6+z$. For a highly biased sample, $a^{\rm (2D)}_{\rm RSD}$ is closer to unity and the redshift evolution is less pronounced. However, the evolution of $a^{\rm (2D)}_{\rm RSD}$ depends on the redshift dependence of the bias; if $b(z)$ is steeply dependent upon redshift then $a^{\rm (2D)}_{\rm RSD}$ will increase with $z$, whereas for a weakly evolving $b(z)$, $a^{\rm (2D)}_{\rm RSD}$ will remain approximately constant or decrease. 

As we are using mock galaxy data, we can calculate the genus amplitude of the density field in both real and redshift space (we define these quantities as $A_{\rm real}$ and $A_{\rm rsd}$ respectively). Since $a_{\rm RSD}^{\rm (2D)}$ is simply the ratio of these quantities we can reconstruct this function using our data. The lightcone data is initially constructed in real space, to generate shells in redshift space we adjust the positions of the mock galaxies $x$ along the line of sight according to 

\begin{equation} x' = x + {v (1+z) \over H(z)} \end{equation} 

\noindent where $v$ is the velocity along the line of sight and $z$ is the redshift of the galaxy. After adjusting the galaxy positions, we bin them into shells as before, Gaussian smooth the field perpendicular to the line of sight then calculate the genus and then the genus amplitude by numerically integrating the curve. 

In the bottom panel of Figure \ref{fig:rsd2} we exhibit the ratio of genus amplitudes $A_{\rm rsd}/A_{\rm real}$ for our mock galaxy data. The green points are snapshot data and yellow stars are the lightcone data shells. The black solid line is the linear theory prediction for $a_{\rm RSD}^{\rm (2D)}$, assuming a bias model of $b(z) = 1.6 + z$, which provides a good fit to the sample under consideration. We find that the linear prediction broadly agrees with the measurement, although the measured ratio is systematically lower by $\sim 1\%$. Similar behaviour can be found in \citet{Appleby:2017ahh}, where the linear RSD prediction was found to be accurate to $\sim 1\%$ for the smoothing scales adopted. 

It is clear that the two dimensional genus amplitude of the redshift space distorted field is significantly altered from its real space counterpart. In a realistic scenario one must constrain the cosmological parameters $\Omega_{\rm mat}$, $w_{\rm de}$ and $\beta$ simultaneously, or otherwise first determine $\beta$ using an independent method. If we know the bias of our galaxy sample, we can correct for the linear RSD effect to $\sim 1\%$. Alternatively, as we will discuss in section \ref{sec:amp}, it is possible to use the dependence of the genus on $\beta$ to directly constrain this parameter. For now we assume that the bias can be accurately measured.

\subsection{Small Scale Systematics}
\label{sec:sss}

As discussed in \citet{Appleby:2017ahh} and also demonstrated in Figures \ref{fig:chi1}, \ref{fig:lc}, the genus amplitude of the non-linear, low-redshift matter density field differs from its Gaussian expectation value. We have argued that redshift space distortion affects this statistic, but even in real space we can state that the genus amplitude $A$ is different from its Gaussian counterpart $A_{\rm G}$. There are a number of reasons for this discrepancy. 

The genus amplitude is given by a ratio of cumulants of the field - to leading order the expression is given by equation ($\ref{eq:gg2d}$). The cumulants themselves are integrals over the entire power spectrum; it follows that they will be affected by the shape of $P(k)$ in the high-$k$ regime. The shot noise contribution modifies the shape of the power spectrum on small scales. Since the galaxy power spectrum will also depend on a multiplicative tracer bias, the integrals in equations ($\ref{eq:s02},\ref{eq:s12}$) will depend on some non-trivial combination of a potentially scale dependent bias and shot-noise. However, this effect can be minimized in a number of ways. One is to ensure that the number density of the galaxy sample is constant in redshift - and the effect on the genus is smaller if we make a mass cut as opposed to randomly selecting galaxies. Clearly the higher the density of the sample the smaller the effect. Finally, the effect of the high-$k$ regime on the integrals in equations ($\ref{eq:s02},\ref{eq:s12}$) is exponentially suppressed by our choice of Gaussian smoothing $R_{\rm G}$ - the larger the smoothing scale the smaller the effect. We note that the effect of shot noise will increase the genus amplitude relative to its Gaussian form \citep{Kim:2014axe,Appleby:2017ahh}.

A second effect is caused by non-linear gravitational collapse, which induces higher order cumulant contributions to the genus. To linear order in $\sigma_{0}$, one can expand the genus in terms of two and three points functions (cf equation ($\ref{eq:mat1}$)). At ${\cal O}(\sigma_{0})$, gravitational collapse does not affect the amplitude $A$. However, we can expect higher order contributions $\sigma_{0}^{n}$, $n>1$ to also be induced, and these will generically modify the amplitude (this effect is so-called `gravitational smoothing' \citep{1989ApJ...345..618M}). This modification to $A$ is suppressed as we increase the smoothing scale $R_{\rm G}$, but is at the percent level for $R_{\rm G} \sim 10 {\rm Mpc}/h$.

\begin{figure}
  \includegraphics[width=0.45\textwidth]{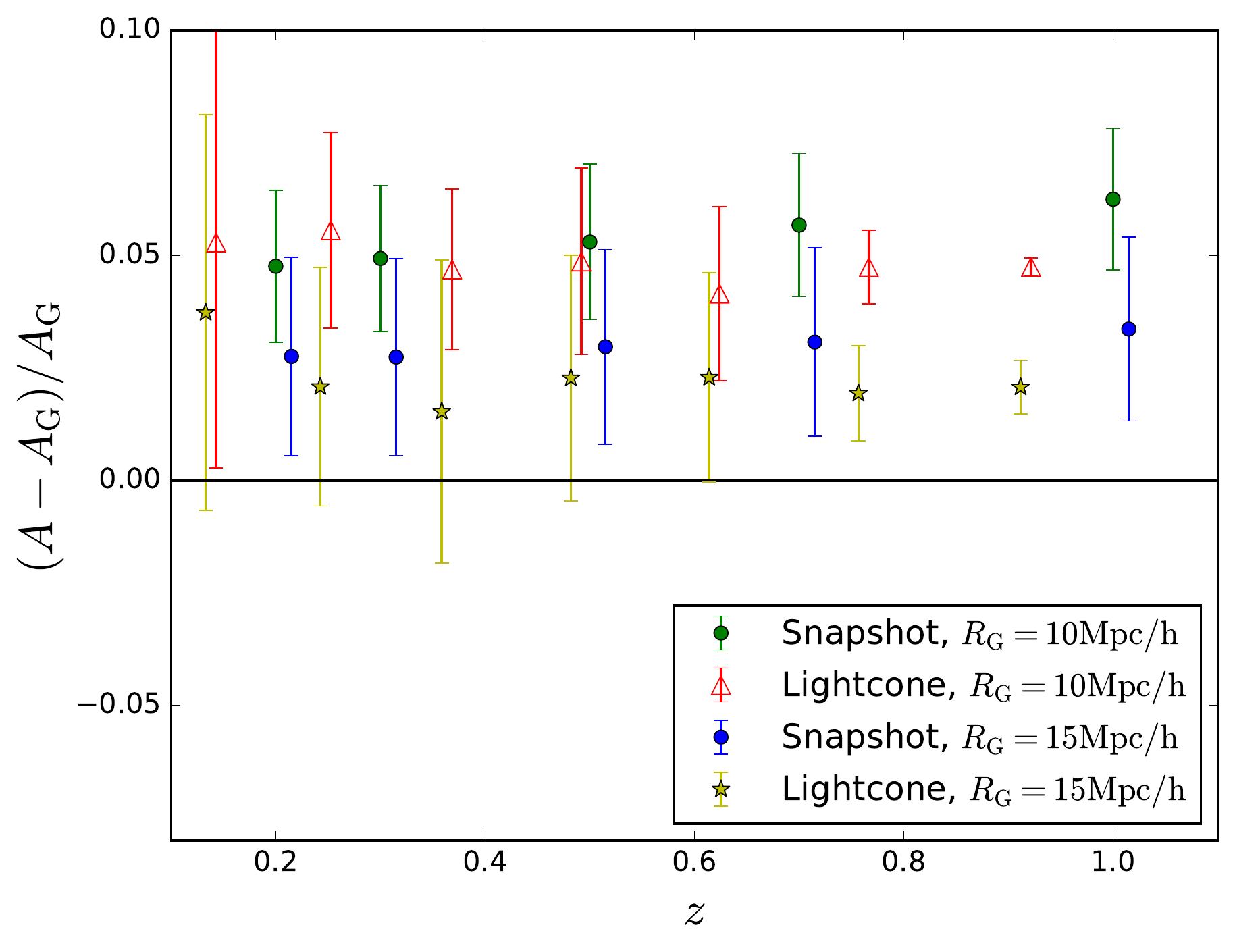}
  \caption{The fractional deviation of the measured genus amplitude relative to the Gaussian expectation value as a function of redshift for snapshot data slices (points/error bars) and lightcone shells (red triangles/yellow stars), taking two smoothing scales $R_{\rm G} = 10 {\rm Mpc}/h$ (red triangles, green points) and $R_{\rm G} = 15 {\rm Mpc}/h$ (yellow stars/blue points). For $R_{\rm G} = 10 {\rm Mpc}/h$, the departure of the measured genus relative to its Gaussian form is $\sim 5\%$, and the snapshot data (green points) show percent level evolution over the range $0.2 < z < 1$. Smoothing over larger scales $R_{\rm G} = 15 {\rm Mpc}/h$ suppresses both effects. }
  \label{fig:shot_noise}
\end{figure}

In Figure \ref{fig:shot_noise} we show the fractional deviation of the measured genus amplitude $A$ relative to the Gaussian expectation value $A_{\rm G}$ as a function of redshift for our mock galaxy data. The green/blue points and error bars are the Horizon Run 4 snapshot data, taking different smoothing scales $R_{\rm G} = 10 {\rm Mpc}/h$ (green) and $R_{\rm G} = 15 {\rm Mpc}/h$ (blue). The red triangles/yellow stars are the genus amplitude measurements for lightcone data with the same smoothing scales. The snapshot and lightcone data are consistent for both smoothing scales. 

For $R_{\rm G} = 10 {\rm Mpc}/h$, the combined `small-scale' behaviour of shot noise and non-linear gravitational collapse produce a  $\sim 5\%$ departure of the measured genus from its Gaussian counterpart, but for our fiducial choice of smoothing scale $R_{\rm G} = 15 {\rm Mpc}/h$ the effect is $\sim 3\%$ and can be reduced further by either increasing $R_{\rm G}$ or $\bar{n}$. Furthermore, for $R_{\rm G} = 15 {\rm Mpc}/h$ the redshift dependence of this effect is negligible $< 1\%$. For $R_{\rm G} = 10 {\rm Mpc}/h$ one can start to observe a stronger (but still marginal) evolution effect, with the amplitude $A$ decreasing to the present.

To account for RSD and small scale systematics, we modify the $\chi^{2}$ distributions that we wish to minimize from their schematic forms ($\ref{eq:chi1},\ref{eq:ch}$) to

\begin{widetext}
\begin{eqnarray} \label{eq:ch1} & &  \chi_{\rm evo}^{2} = \sum_{i} {\left( A_{i}(z_{i},\Omega_{\rm mat},w_{\rm de})/a_{\rm RSD}^{\rm (2D)} - A_{0} \right)^{2} \over \sigma_{i}^{2} } \\ 
\label{eq:ch2} & &  \chi_{\rm mag}^{2} = \sum_{i} {\left( A_{i}(z_{i},\Omega_{\rm mat},w_{\rm de})(1-\Delta_{\rm SN})/a_{\rm RSD}^{\rm (2D)} - A_{\rm G}(\Omega_{\rm mat}) \right)^{2} \over \sigma_{i}^{2} + \sigma_{\rm RSD}^{2} + \sigma_{\rm SN}^{2}} \end{eqnarray} 
\end{widetext}

\noindent where $\sigma_{\rm RSD}$ and $\sigma_{\rm SN}$ are uncertainties associated with the RSD and small scale modifications to the amplitude and $\Delta_{\rm SN}$ is a systematic correction due to shot noise. These quantities are estimated from snapshot mock galaxy catalogs. We multiply the measured amplitudes $A_{i}$ by $[a_{\rm RSD}^{\rm (2D)}]^{-1}$ to eliminate the effect of linear RSD; we have argued that this correction is accurate to $\sim 1\%$, so take $\sigma_{\rm RSD} = 0.01 A_{\rm G}$ in what follows.

The correction factor $\Delta_{\rm SN}$ in equation ($\ref{eq:ch2}$) is introduced to eliminate the known systematic of shot noise which increases the measured amplitude relative to its Gaussian expectation value. Its value is estimated from the mock snapshot data - for $R_{\rm G} = 15 {\rm Mpc}/h$ we take a constant $\Delta_{\rm SN} = 0.025 A_{\rm G}$ over the redshift range $0.1 < z < 1$.

The uncertainty $\sigma_{\rm SN}$ is the statistical uncertainty associated with $\Delta_{\rm SN}$ and is also obtained from the snapshot data - for the $3150\times 3150 ({\rm Mpc}/h)^{2}$ slices we find a statistical uncertainty of $\sigma_{\rm SN} \sim 2\%$ for our fiducial parameter set $\Delta = 60 {\rm Mpc}/h$, $R_{\rm G} = 15 {\rm Mpc}/h$ and $\bar{n} = 10^{-3} ({\rm Mpc}/h)^{-3}$. We therefore take

\begin{equation} \sigma_{\rm SN} = 0.02 \sqrt{3150^{2} \over 4\pi d^{2}_{\rm cm}(z_{\rm i})} A_{\rm G} \end{equation} 

\noindent for our lightcone data, where $d_{\rm cm}(z_{\rm i})$ is the comoving distance to the $i^{\rm th}$ redshift shell. 

It is important to stress that by correcting the observed amplitudes by $\Delta_{\rm SN}$, we have made an implicit model dependent assumption that the actual galaxy data is accurately represented by the mock catalogs from which we estimate this quantity. Although the correction $\Delta_{\rm SN} \sim 3\%$ is small, any discrepancy between the mock catalogs and data will bias our resulting cosmological parameter reconstruction. When applying the statistics to data, one should check carefully that no bias is introduced by repeating the analysis for various $R_{\rm G}$ smoothing scales. 

We do not add the contributions $\sigma_{\rm RSD}$, $\Delta_{\rm SN}$ and $\sigma_{\rm SN}$ to $\chi_{\rm evo}^{2}$, as these effects do not modify the evolution of the genus amplitude, only its magnitude. The redshift evolution associated with RSD is reproduced well by the correction $a_{\rm RSD}^{\rm (2D)}$, and the redshift evolution of the shot noise correction is negligible if we impose a constant galaxy density cut.

\section{Constraints on $(\Omega_{\rm \MakeLowercase{mat}}, \MakeLowercase{w_{\rm de}})$ from the Genus Amplitude}
\label{sec:ampr}

We now consider the parameter constraints that can be placed on $\Omega_{\rm mat}, w_{\rm de}$ by minimizing the two distributions ($\ref{eq:ch1},\ref{eq:ch2}$). For each set of parameters $(\Omega_{\rm mat}, w_{\rm de})$ we calculate the genus in the $N_{\rm shell} = 36$ redshift shells, and extract the amplitude. We construct the mean $\bar{A}_{i}$ and error on the mean $\bar{\sigma}_{i}$ of each set of five adjacent shells, and use these as $A_{i}(z_{i},\Omega_{\rm mat},w_{\rm de})$ and $\sigma_{i}$ in the $\chi^{2}$ distributions. We adopt $b(z) = 1.6 + z$ and correct the observed amplitude by $a_{\rm RSD}^{\rm (2D)}(z_{i},\Omega_{\rm mat},w_{\rm de},b(z))$, which also depends on the cosmological model adopted. We then fit a constant $A_{0}$ to each set of measurements in the case of minimizing $\chi_{\rm evo}^{2}$, or compare the measurements to the Gaussian expectation $A_{\rm G}(\Omega_{\rm mat})$ when using $\chi_{\rm mag}^{2}$. Our method of reconstructing $A_{\rm G}(\Omega_{\rm mat})$ was described in section \ref{sec:1}.

\begin{figure}
  \includegraphics[width=0.45\textwidth]{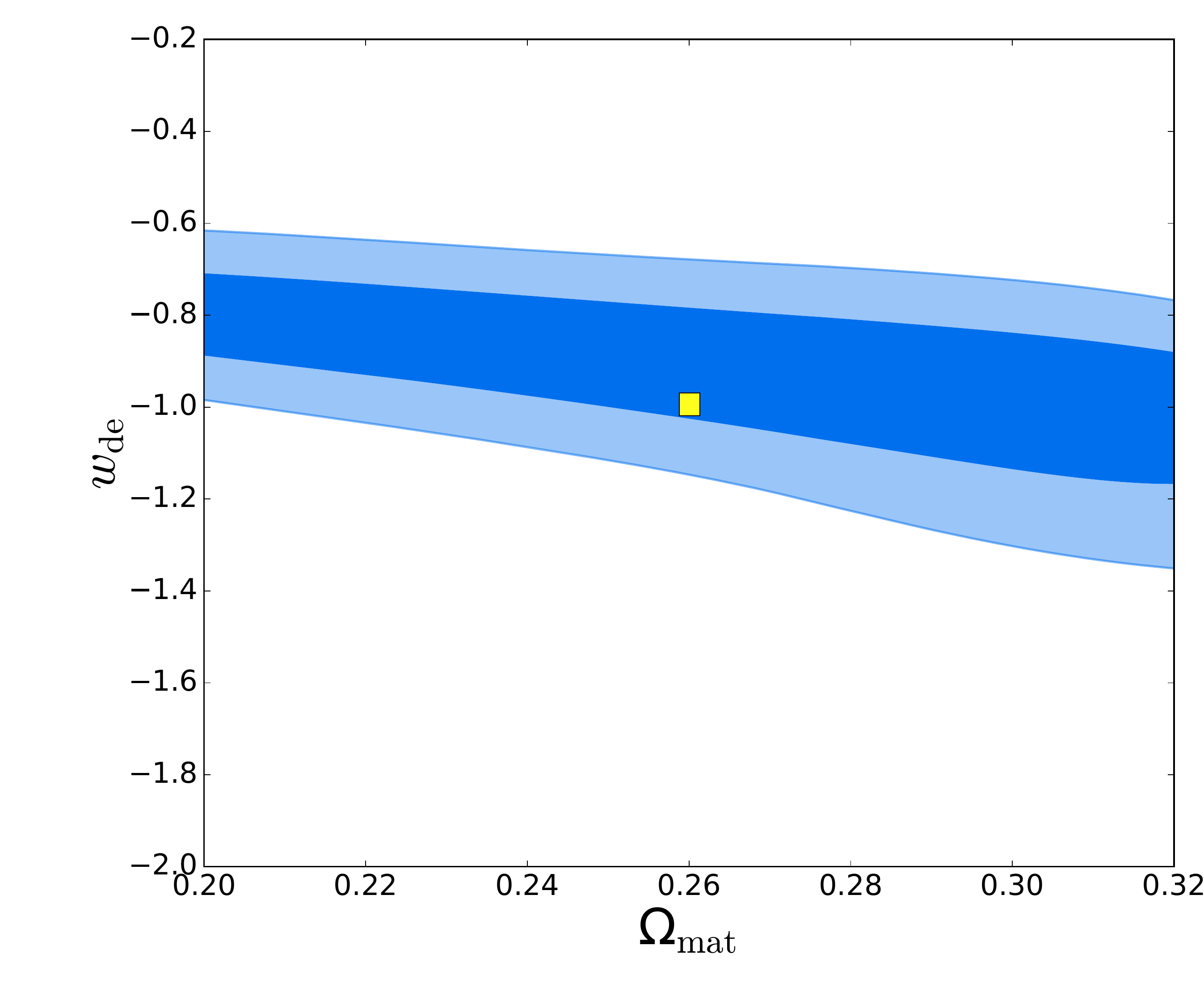}\\
  \includegraphics[width=0.45\textwidth]{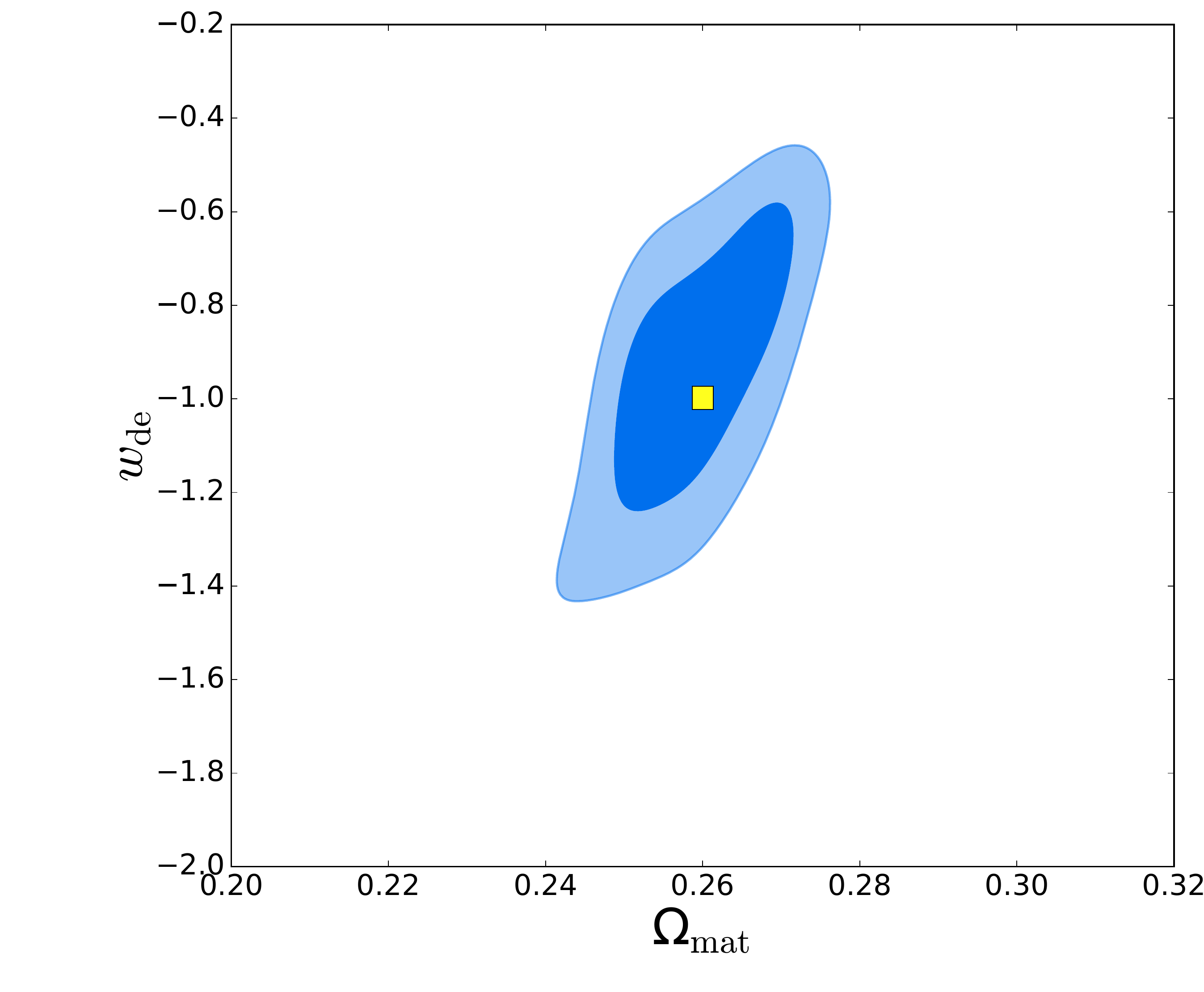}\\
  \caption{Two dimensional posterior probability distributions when we use cosmological parameters $(\Omega_{\rm mat},w_{\rm de})$ to infer the distance redshift relation and minimize the $\chi^{2}$ distribution $\chi_{\rm evo}^{2}$ (top panel) and $\chi_{\rm mag}^{2}$ (bottom panel). Dark/light blue contours indicate the $1,2-\sigma$ parameter regions. The input cosmology, denoted as a yellow square, is reproduced in both approaches.}
  \label{fig:contour_real}
\end{figure}

We perform a Monte Carlo Markov Chain exploration of the two dimensional parameter space, using the publicly available software ${\rm CosmoMC}$ \citep{Lewis:2013hha,Lewis:2002ah} as a generic sampler to minimize the $\chi^{2}$ distribution and generate the posterior probability contours of $(\Omega_{\rm mat},w_{\rm de})$. We exhibit the two dimensional $68\%$ and $95\%$ confidence intervals in Figure \ref{fig:contour_real}, minimizing $\chi_{\rm evo}^{2}$ (top panel) and $\chi_{\rm mag}^{2}$ (bottom panel). We have fixed all other cosmological parameters to their WMAP5 values, and $h=0.72$. We do not expect this statistic to be sensitive to the primordial amplitude $A_{\rm s}$ (as we are measuring the ratio of cumulants) or tilt $n_{\rm s}$ given the extremely tight CMB constraints on this quantity. The solid yellow squares in the figures represent the fiducial cosmology, which lie within the 1--$\sigma$ limits. The one dimensional marginalised parameter constraints are given in Table \ref{tab:1}, labeled `Evo' and `Mag' respectively. 

Both methods provide a consistent reconstruction of the input cosmological model. If we use the evolution of the amplitude only, then we arrive at a constraint of $\Delta w \sim 0.15$ on the equation of state of dark energy but an extremely weak constraint on the matter density. We could anticipate such a result from the bottom panel of Figure \ref{fig:2}. The slight degeneracy between $w_{\rm de}$ and $\Omega_{\rm mat}$ is due to the manner in which these two parameters modify $A$ - increasing $w_{\rm de}$ and $\Omega_{\rm mat}$ will have similar effects on the redshift dependence of $A$. The data exhibits a slight bias towards a higher value of $w_{\rm de} > -1$, although the effect is not significant. This is due to the slightly high value of the genus amplitude measurement in the lowest redshift bin.

If we directly compare the measured genus amplitude to its Gaussian expectation value and minimize $\chi_{\rm mag}^{2}$ (bottom panel), then we significantly tighten the constraint on $\Omega_{\rm mat}$. The cumulants of the field (which comprise the amplitude $A$) are sensitive to the peak of the power spectrum, which in turn depends on $\Omega_{\rm mat}$. As we increase $\Omega_{\rm mat}$, $k_{\rm eq}$ also increases and generates more power at smaller scales, which produces a larger genus amplitude (and vice versa if we decrease $\Omega_{\rm mat}$). The parameter degeneracy between $\Omega_{\rm mat}$ and $w_{\rm de}$ has now reversed compared to the top panel. The reason for this effect is that the best measured values of $A$ are the high redshift shells, so in the language of Figure \ref{fig:2} the dominant constraining power arises from the high-$z$ behaviour of the top and bottom panels. Incorrectly assuming a less negative equation of state (red line, bottom panel) is mildly degenerate with selecting a high $\Omega_{\rm mat}$ value (blue dashed line in the top panel) and vice versa. 

The conclusion that we arrive at this point is that both methods will place comparable, but relatively modest constraints on the equation of state of dark energy. This constraint can be improved by reducing $R_{\rm G}$ or by measuring the genus at higher redshift. However, as we decrease $R_{\rm G}$ one can expect the genus amplitude to evolve due to non-linear gravitational collapse and one must model these higher order corrections. The advantage of using this statistic is that the non-linear correction is small in the mildly non-linear regime; at scales $R_{\rm G} \sim 10 {\rm Mpc}/h$ the evolution of the amplitude is a $\sim 1\%$ effect.  

Both methods of cosmological parameter estimation require knowledge of the bias of the galaxy sample being used, and when minimizing $\chi^{2}_{\rm mag}$ we also require an estimate of $\Delta_{\rm SN}$ and $\sigma_{\rm SN}$. These quantities can be estimated with mock galaxy catalogs, although the bias must be independently ascertained from the data. 

The sensitivity of the genus curve to $\beta$ suggests that measuring the statistic in redshift space will allow us to place constraints on both $\beta$ and $\Omega_{\rm mat}$.  As we are only using the genus amplitude in this work, the effect of $\beta$ and $\Omega_{\rm mat}$ will be degenerate, but as we will now show it is possible to break this degeneracy by using three dimensional information of the density field.

\begin{table}
\begin{center}
 \begin{tabular}{||c | c | c  ||}
 \hline
 Parameter & Evo &  Mag  \\ [0.5ex] 
 \hline\hline
  $\Omega_{\rm mat}$ & $0.262^{+0.081}_{-0.032}$ & $0.259^{+0.011}_{-0.009}$   \\ 
 \hline
  $w_{\rm de}$ & $-0.92^{+0.17}_{-0.11}$ & $-0.94^{+0.20}_{-0.21}$   \\ 
 \hline
\end{tabular}
\caption{\label{tab:1}  Marginalised one dimensional parameter constraints on $(\Omega_{\rm mat}, w_{\rm de})$ using the evolution of the genus amplitude only (column `Evo') and using the magnitude (column `Mag'). }
\end{center} 
\end{table}

\section{Breaking the $\beta - \Omega_{\rm \lowercase{mat}}$ degeneracy using a combination of two and three-dimensional genus measurements}
\label{sec:amp}

As discussed in the previous section, the two dimensional genus amplitude in redshift space is significantly affected by peculiar velocities and also tracer bias. The effect of varying $\Omega_{\rm mat}$ on the genus amplitude will be degenerate with the redshift space distortion parameter $\beta = f/b$. 

We can break this degeneracy by noting that the effect of redshift space distortion on the three dimensional genus amplitude is different to two dimensional slices. Hence by measuring the genus of both two dimensional slices and the full three dimensional field, we can constrain the RSD parameter $\beta$ and $\Omega_{\rm mat}$ simultaneously. 

The genus for a three dimensional Gaussian random field is given by equation ($\ref{eq:gg}$). To leading order in a perturbative expansion of the density perturbation, the three dimensional genus in real and redshift space are related via an amplitude change \citep{1996ApJ...457...13M, Matsubara:2000mw, Codis:2013exa, Kim:2014axe} - 

\begin{eqnarray} \label{eq:3d_rsd} & & g_{\rm 3D}^{\rm rsd}(\nu) = a_{\rm RSD}^{\rm (3D)}  g_{\rm 3D}^{\rm real}(\nu) , \end{eqnarray} 

\noindent where 

\begin{equation}\label{eq:amp3d_rsd} a_{\rm RSD}^{\rm (3D)} = \sqrt{3}{3 \over 2} \sqrt{C_{1} \over C_{0}} \left( 1 - {C_{1} \over C_{0}}\right) , \end{equation}

\begin{figure}
  \includegraphics[width=0.45\textwidth]{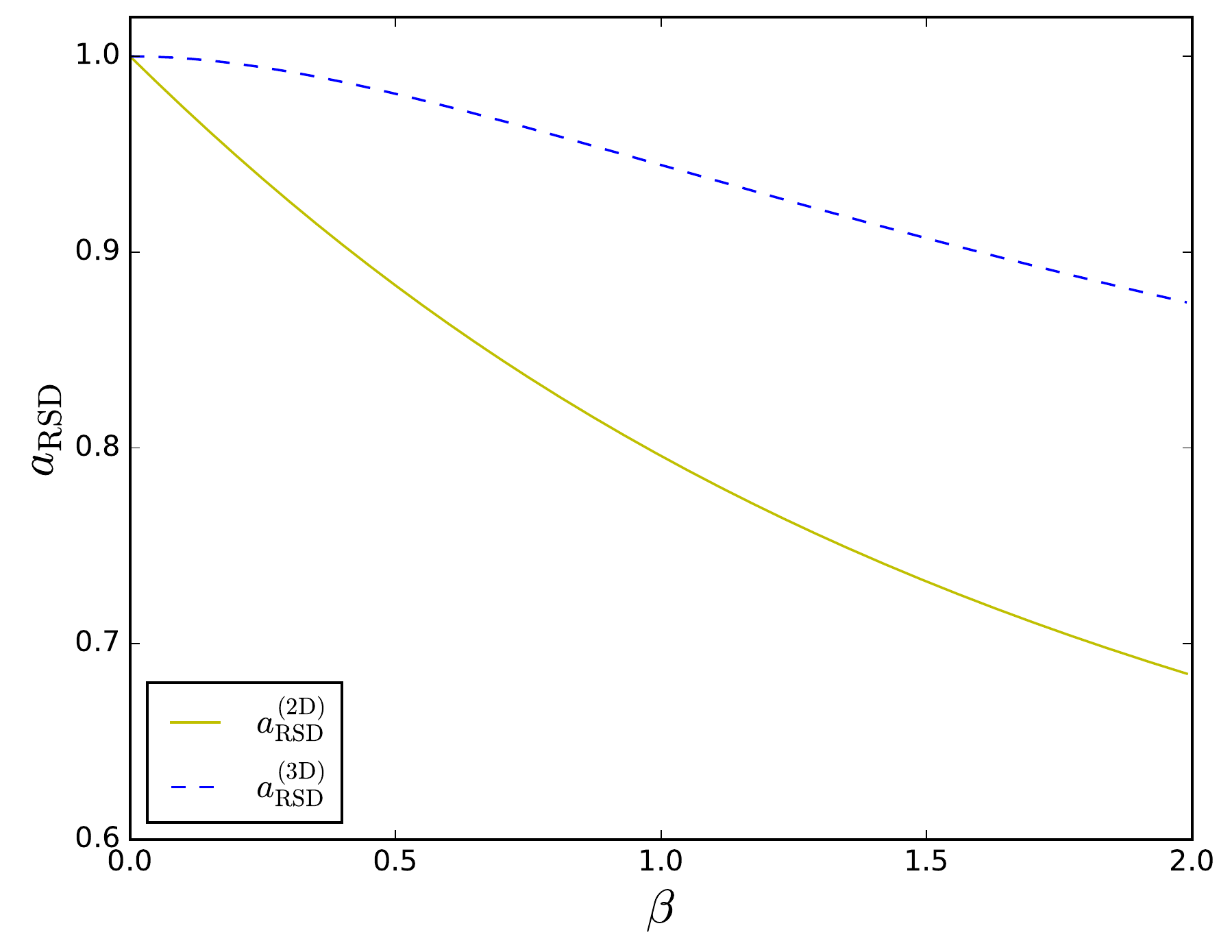}\\
  \caption{The fractional change in the genus amplitude in redshift space relative to its real space counterpart, as a function of the redshift space distortion parameter $\beta$. The two and three dimensional genus amplitudes, exhibited as yellow solid and blue dashed lines respectively, exhibit different sensitivity to this parameter. This allows us to break the degeneracy between $\Omega_{\rm mat}$ and $\beta$ that would otherwise be present if only using $g_{\rm 2D}$.}
  \label{fig:rsd3}
\end{figure}

\noindent The dependence of $a_{\rm RSD}^{\rm (3D)}$ and $a_{\rm RSD}^{\rm (2D)}$ on $\beta$ are exhibited in Figure \ref{fig:rsd3}. It is clear that the effect of linear redshift space distortion on the three dimensional genus is significantly lower than in the two dimensional case. Over the redshift range $0 < z < 1$, the evolution of $a_{\rm RSD}^{\rm (3D)}$ is typically $< 2\%$ for $\beta < 0.4$.

The fact that the two and three dimensional genus measurements present different sensitivity to the growth factor allows us to break the degeneracy between $f/b$ and $\Omega_{\rm mat}$. Roughly speaking, the three dimensional genus amplitude is insensitive to $f/b$ and provides a measurement of $\Omega_{\rm mat}$, and the two dimensional genus can then be used to constrain $\beta$. The constraints are model dependent, and valid in the regime in which the linear redshift space distortion effect is applicable.

We apply our test to a mock data sample. Of our seven measured two dimensional genus amplitudes $\bar{A}_{i}$, we use the two lowest and two highest redshift measurements, which means that we use all redshift shells in the range $0.1 < z < 0.308$ and $0.75 < z < 1.03$. Instead of using the intermediate redshift shells, we calculate the three dimensional genus of the density field in a $V = 1 \left({\rm Gpc/h}\right)^{3}$ volume at $z=0.5$. For realistic galaxy data the observed volume would have an irregular shape and be subject to masks and boundaries, but for simplicity we take a cube. We assume that the two and three dimensional samples are sufficiently distant such that they can be considered uncorrelated measurements.

The genus amplitude of the full three dimensional Horizon Run snapshot box is calculated at $z=0.5$, using a three dimensional Gaussian smoothing scale $\Lambda_{\rm G}  = 20 {\rm Mpc}/h$. The mock galaxy positions are adjusted to account for redshift space distortions. We generate $N_{\rm mock} = 500$ mock Gaussian density fields drawn from the same underlying power spectrum as the one used in the Horizon Run 4 simulation, in a $1 ({\rm Gpc/h})^{3}$ volume. We calculate the mean and rms fluctuations of the genus for this fiducial Gaussian field, finding a statistical uncertainty on the amplitude of $\sim 3\%$ for this volume. Therefore in what follows we take $\sigma^{\rm (3D)}_{z=0.5} = 0.03 A^{\rm (3D)}_{z=0.5}$, where $A^{\rm (3D)}_{z=0.5}$ is the three dimensional genus amplitude measured from the Horizon Run simulation volume. 

We measure $g_{\rm 3D}$ over $N=100$ values of $\nu_{\rm A}$, equi-spaced over the range $-4 < \nu_{\rm A} < 4$. Similarly to the two dimensional case, we calculate the amplitude of the genus from the $g_{\rm 3D}(\nu_{\rm A})$ curve by integrating over $\nu_{\rm A}$ with window function $H_{2}(\nu_{\rm A})$. The result is the genus amplitude multiplied by a factor of $2\sqrt{2\pi}$.

We are using a simulated data set of fixed comoving volume at $z=0.5$. When measuring the 3D genus of the field using assumed parameters $(\Omega_{\rm mat}, w_{\rm de}) \ne (\Omega_{\rm mat}^{\rm (fid)}, w_{\rm de}^{\rm (fid)})$ we must perform a similar adjustment as in the two dimensional case -  we adjust the smoothing scale and volume of the three dimensional genus according to 

\begin{equation} g_{\rm 3D} = {G_{\rm 3D}[\Lambda'_{\rm G}] \over V_{\rm fid}} {\alpha_{\rm X} \over \alpha_{Y}} \end{equation}

\noindent where $G_{\rm 3D}$ is the dimensionless genus of the three dimensional field, $V_{\rm fid}$ is the fiducial snapshot box volume and 

\begin{equation} {\alpha_{X} \over \alpha_{Y}} = {D_{\rm A, X}^{2}(z) H_{Y}(z) \over D_{\rm A, Y}^{2}(z) H_{X}(z)} \end{equation} 

\noindent where $D_{\rm A, X,Y}$, $H_{X,Y}$ are the angular diameter distance and Hubble parameter for the fiducial ($X$) and assumed ($Y$) cosmological model. The fiducial smoothing scale is $\Lambda_{\rm G} = 20 {\rm Mpc}/h$ and $\Lambda'_{\rm G} = \left[\alpha_{\rm X}/\alpha_{\rm Y}\right]^{1/3}\Lambda_{\rm G}$. 

We minimize the following $\chi^{2}$ distribution 

\begin{widetext}
\begin{equation}\label{eq:chifb} \chi^{2} = \sum_{i} {\left( A_{i}(z_{i},\Omega_{\rm mat},w_{\rm de})(1-\Delta_{\rm SN}) [a_{\rm RSD}^{\rm (2D)}(\beta_{i})]_{z_{i}}^{-1} - A_{\rm G}^{\rm (2D)}(\Omega_{\rm mat},w_{\rm de}) \right)^{2} \over \sigma_{i}^{2} + \sigma_{\rm RSD}^{2} + \sigma_{\rm SN}^{2}} + {\left[ A^{\rm (3D)}_{z=0.5}[a_{\rm RSD}^{\rm (3D)}(\beta_{z=0.5})]^{-1} - A^{\rm (3D)}_{\rm G}(\Omega_{\rm mat},w_{\rm de}) \right]^{2} \over \sigma_{z=0.5}^{2}} \end{equation} 
\end{widetext}

\noindent where $A_{\rm G}^{\rm (2D)}(\Omega_{\rm mat})$ and $A^{\rm (3D)}_{\rm G}(\Omega_{\rm mat})$ are the two and three dimensional Gaussian amplitudes, and we have assumed that $\beta_{i}$ are independent and arbitrary free parameters at each redshift. The $i^{\rm th}$ redshift shell depends upon the value of $[a_{\rm RSD}^{\rm (2D)}(\beta_{i})]$, which is a function only of $\beta_{i} = \beta(z_{i})$. The $i$ subscript runs over $i=1,2,6,7$, denoting the two dimensional measurements that we use in our analysis. We vary the parameters $\Omega_{\rm mat}, w_{\rm de}$ and $\beta_{1,2,6,7}$. After initially also varying the parameter $\beta_{z=0.5}$, we found our posterior distributions to be completely insensitive to its value - this is a consequence of the three dimensional genus amplitude being insensitive to RSD effects. We therefore fix $\beta(z=0.5)$ to a fiducial value $\beta(z=0.5)=0.3$ in what follows, with the understanding that our result is insensitive to this parameter.

\begin{figure*}
  \includegraphics[width=0.95\textwidth]{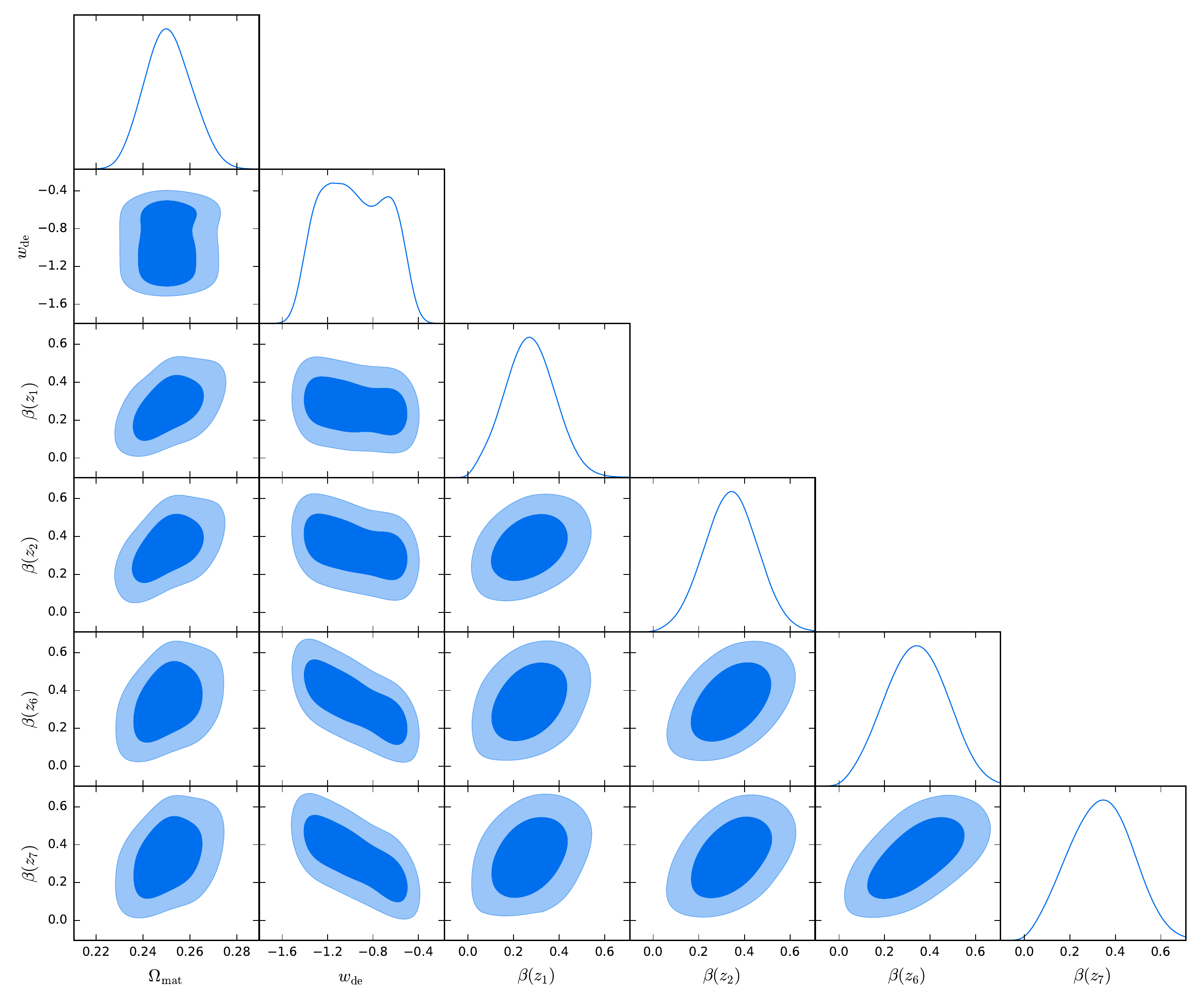}
  \caption{Triangle plot of the parameters $\beta_{1,2,6,7}$, $w_{\rm de}$ and $\Omega_{\rm mat}$ using four measurements of the two dimensional genus and a measurement of the three dimensional genus amplitude at $z=0.5$. A single measurement of the three dimensional genus is sufficient to eliminate the degeneracy between $\Omega_{\rm mat}$ and $\beta_{i}$, allowing joint constraints to be made.}
  \label{fig:triangle}
\end{figure*}

We minimize the distribution ($\ref{eq:chifb}$) and exhibit the resulting two dimensional marginalised constraints in Figure \ref{fig:triangle}. There is now effectively no constraint on $w_{\rm de}$ - an additional degeneracy is introduced between $\beta_{i}$ and $w_{\rm de}$, which further restricts our ability to constrain the equation of state of dark energy. However, we obtain competitive constraints on both $\Omega_{\rm mat}$ and $\beta_{i}$ -  the one dimensional parameter constraints are shown in Table \ref{tab:tri} - the constraint on $\Omega_{\rm mat}$ is $\Delta \Omega_{\rm mat} \sim 0.01$ and we find a simultaneous constraint of order $\sim 0.1$ on $\beta$ at each redshift.

\begin{figure}
  \includegraphics[width=0.45\textwidth]{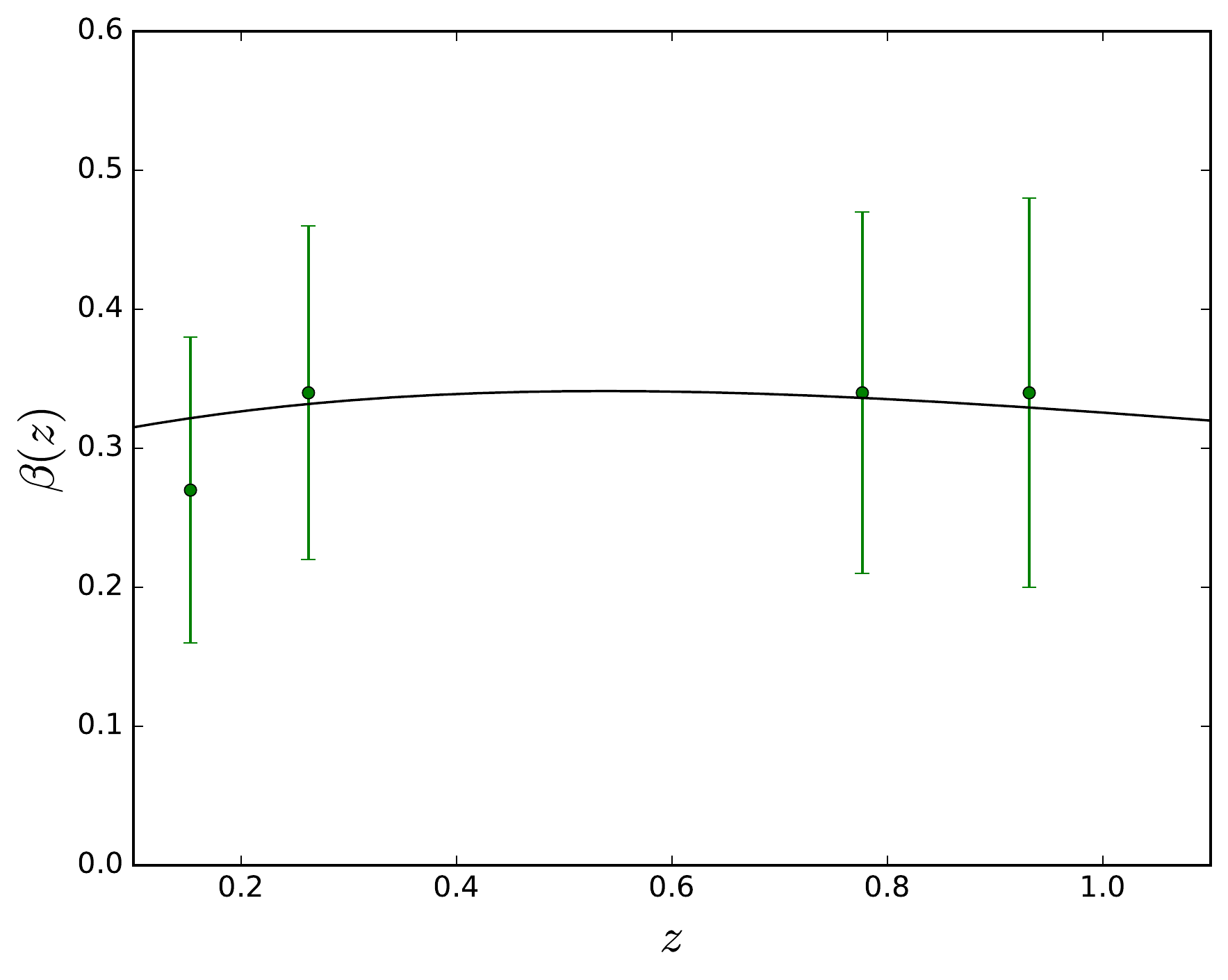}
  \caption{Marginalised parameter constraints on $\beta(z)$ based on four measurements of the two dimensional genus amplitude using mock lightcone data (green points/error bars). The fiducial input value of $\beta(z) \simeq \Omega_{\rm m}^{\gamma}(z)/b(z)$ is exhibited as a solid black line, where we have used $\gamma = 6/11$, $\Omega_{\rm mat} = 0.26$ and $b(z) = 1.6 + z$. The reconstruction is consistent with the input model.}
  \label{fig:bet}
\end{figure}

Finally, we exhibit the best fit reconstructed $\beta_{i}$ measurements and the underlying theoretical expectation value $\beta(z) = \Omega_{\rm m}^{\gamma}(z)/b(z)$ for our data set in Figure \ref{fig:bet}. We have used the fiducial cosmology of the Horizon Run 4 simulation and our knowledge of the mock galaxy bias $b(z)=1.6 +z$ to generate the solid black line. We set $\gamma = 6/11$, as per the standard $\Lambda$CDM expectation value. Our analysis correctly reproduces the fiducial cosmology $\Omega_{\rm mat}$ and the growth rate $\beta(z)$. There is a small but statistically insignificant bias towards a lower $\Omega_{\rm mat}$ and $\beta(z)$ relative to the input cosmology in the lowest redshift shell, due to the slightly high value of the genus in this bin.

\begin{table}
\begin{center}
 \begin{tabular}{||c | c ||}
 \hline
 Parameter &  68\% limits, 2D \& 3D \\ [0.5ex] 
 \hline\hline
$\Omega_{\rm mat}$ &  $0.251\pm 0.01$ \\
 \hline
$w_{\rm de}$ &  $-1.03^{+0.28}_{-0.24}$ \\
 \hline
$\beta (z_{1})$ &  $0.27^{+0.10}_{-0.11}$ \\
 \hline
$\beta (z_{2})$ &  $0.34\pm 0.11$ \\
 \hline
$\beta (z_{6})$ &  $0.34\pm 0.12$ \\
 \hline
$\beta (z_{7})$ &  $0.34\pm 0.13$ \\
 \hline
\end{tabular}
\caption{\label{tab:tri} One dimensional marginalised parameter constraints on $\Omega_{\rm mat}$ and $\beta(z_{\rm i})$ using a combination of four, two-dimensional genus measurements and a three dimensional genus measurement at $z=0.5$. The three dimensional genus breaks the degeneracy between $\Omega_{\rm mat}$ and $\beta$ in redshift space, allowing us to place joint parameter constraints. }
\end{center} 
\end{table}

\section{Discussion}

In this work we have studied the genus amplitude of two dimensional slices of the three dimensional dark matter density field, and considered how this quantity can be used for parameter estimation. There are numerous advantages to using this statistic - to leading order in a $\sigma_{0}$ expansion it is unaffected by non-linear gravitational collapse. Furthermore, in principle it is independent of the growth rate and linear bias of the galaxy sample. Selecting thick two dimensional slices can largely eliminate non-linear Finger of God effects, which are generated from highly non-linear processes and would otherwise have to be modeled. In essence, the statistic is a measure of the number of structures at a given smoothing scale. It depends on the shape of the matter density power spectrum in the case of a Gaussian density field. If we select an `incorrect' cosmological parameter set to infer the distance-redshift relation, we also expect a systematic shift in the genus amplitude with redshift. This introduces a parameter degeneracy between $\Omega_{\rm mat}$ and $w_{\rm de}$. We have explained how both the redshift evolution and magnitude of the genus contains information regarding the underlying cosmology.

Redshift evolution of the genus is a consequence of selecting the incorrect cosmological model in the distance redshift relation. Constraints on the equation of state of dark energy can be obtained by minimizing the redshift dependence, but this requires a precision measurement $\sim 1 \%$ of the genus amplitude at $z < 0.2$. This will provide the dominant challenge of the method, but such a measurement can be made by utilising all information tomographically. Alternatively one might use the three-dimensional genus amplitude for the low redshift measurement, and combine the result with two dimensional slices at higher redshift. Additional care must be taken for a redshift space distorted density field, as the redshift dependence of the tracers can introduce additional redshift evolution of the statistic. To obtain a constraint on $w_{\rm de}$ using real galaxy data will require accurate knowledge of the linear bias of the sample. The three dimensional genus is less sensitive to RSD effects and is perhaps better suited to a study of this kind. 

The magnitude of the genus amplitude is sensitive to $\Omega_{\rm mat}$ - this is due to the fact that one measures the shape of the linear matter power spectrum via cumulants of the field, and thus the statistic is sensitive to the location of the power spectrum peak. Potentially strong constraints on $\Omega_{\rm mat}$ can be placed, however the two dimensional amplitude is particularly sensitive to both linear redshift space distortions and also a combination of shot noise and non-linear gravitational evolution. The latter is a systematic that can be reduced by smoothing over large scales (and using dense galaxy samples). We have found that the shot noise effect is $\sim 3\%$ when using the smoothing scales and number densities adopted in this work. The linear RSD correction to the genus is sufficient to cancel this effect to $\sim 1\%$ accuracy, but requires knowledge of the bias.

If we do not have an accurate measure of $b(z)$, the effect of RSD is to introduce a parameter degeneracy between $\Omega_{\rm mat}$, $w_{\rm de}$ and the redshift space parameter $\beta(z) = f(z)/b(z)$. This degeneracy can be broken by noting that the two and three dimensional genus amplitudes possess different sensitivity to $\beta(z)$. Hence by measuring combinations of three and two-dimensional data sets, we can provide simultaneous and competitive constraints on $\Omega_{\rm mat}$ and $\beta(z)$. For a single $3\%$ accurate measurement of the three dimensional genus curve, we can place constraints of order $\Omega_{\rm mat} = 0.26 \pm 0.01$. Making subsequent genus measurements of two dimensional slices of the density field at different redshifts allows us to constrain $\beta(z)$ to accuracy $\Delta \beta \simeq 0.1$ at the redshift at which we make the measurement. 

It is important to stress the difference between using the genus amplitude as a cosmological probe compared to more traditional methods such as measuring the power spectrum directly. In fact, when smoothing the density field over sufficiently large scales, measuring the genus amplitude as a function of smoothing scale is equivalent to measuring the shape of the power spectrum over the same scales. However, the advantage of using the genus appears as we try to use the observed data in the mildly non-linear regime. This situation is necessary to maximize the statistical power of the data. For scales corresponding to the non-linear gravitational evolution regime, the shape of the power spectrum distorts considerably and is no longer directly related to the genus amplitude. A large correction to the power spectrum is needed even at $k = 0.1 - 0.2 {\rm h/Mpc}$ to infer its original shape. Conversely the genus, being a topological measure, is unchanged up to the second-order perturbation theory and retains memory of the linear regime even in the quasi-linear regime. Therefore a much smaller correction is needed for non-linear gravitational evolution effects. Hence the genus amplitude is a more conserved quantity against the non-linear gravitational evolution compared to the power spectrum shape.

This fact is crucial for our analysis, as it allows us to use the genus amplitude as a standard population even when we smooth the density field at mildly non-linear scales. It also allows us to compare the observed genus amplitude to the Gaussian expectation value. This property, coupled to the relative insensitivity of the statistic to galaxy bias, are the principle advantages of using the genus over more traditional methods of cosmological parameter estimation.

The next stage of our analysis is the application of the statistics to real galaxy data. This will introduce additional complexities such as masks, boundaries and photometric redshift uncertainties. Such issues will be addressed elsewhere.

\acknowledgements{The authors thank the Korea Institute for Advanced Study for providing computing resources (KIAS Center for Advanced Computation Linux Cluster System) for this work.

This work was supported by the Supercomputing Center/Korea Institute of Science and Technology Information, with supercomputing resources including technical support (KSC-2013-G2-003) and the simulation data were transferred through a high-speed network provided by KREONET/GLORIAD.

Some of the results in this paper have been derived using the HEALPix \citep{Gorski:2004by} package.}

\bibliographystyle{ApJ}
\bibliography{biblio}{}

\begin{thebibliography}{}
\expandafter\ifx\csname natexlab\endcsname\relax\def\natexlab#1{#1}\fi

\bibitem[{Adler(1981)}]{Adler}
Adler, R. 1981, The Geometry of Random Fields (Wiley)

\bibitem[{Anderson {et~al.}(2013)}]{Anderson:2012sa}
Anderson, L., {et~al.} 2013, Mon. Not. Roy. Astron. Soc., 427, 3435

\bibitem[{Appleby {et~al.}(2017)Appleby, Park, Hong, \& Kim}]{Appleby:2017ahh}
Appleby, S., Park, C., Hong, S.~E., \& Kim, J. 2017, Astrophys. J., 836, 45

\bibitem[{Blake {et~al.}(2014)Blake, James, \& Poole}]{Blake:2013noa}
Blake, C., James, J.~B., \& Poole, G.~B. 2014, MNRAS, 437, 2488

\bibitem[{Blake {et~al.}(2011)Blake, Davis, Poole, Parkinson, Brough, Colless,
  Contreras, Couch, Croom, Drinkwater, Forster, Gilbank, Gladders, Glazebrook,
  Jelliffe, Jurek, Li, Madore, Martin, Pimbblet, Pracy, Sharp, Wisnioski,
  Woods, Wyder, \& Yee}]{doi:10.1111/j.1365-2966.2011.19077.x}
Blake, C., Davis, T., Poole, G.~B., {et~al.} 2011, MNRAS, 415, 2892

\bibitem[{Choi {et~al.}(2013)Choi, Kim, Rossi, Kim, \& Lee}]{Choi:2013eej}
Choi, Y.-Y., Kim, J., Rossi, G., Kim, S.~S., \& Lee, J.-E. 2013, Astrophys. J.
  Suppl., 209, 19

\bibitem[{Choi {et~al.}(2010)Choi, Park, Kim, Gott, Weinberg, Vogeley, \&
  Kim}]{Choi:2010sx}
Choi, Y.-Y., Park, C., Kim, J., {et~al.} 2010, ApJS., 190, 181

\bibitem[{Codis {et~al.}(2013)Codis, Pichon, Pogosyan, Bernardeau, \&
  Matsubara}]{Codis:2013exa}
Codis, S., Pichon, C., Pogosyan, D., Bernardeau, F., \& Matsubara, T. 2013,
  Mon. Not. Roy. Astron. Soc., 435, 531

\bibitem[{Cole {et~al.}(2005)}]{Cole:2005sx}
Cole, S., {et~al.} 2005, Mon. Not. Roy. Astron. Soc., 362, 505

\bibitem[{{Coles} {et~al.}(1993){Coles}, {Moscardini}, {Plionis}, {Lucchin},
  {Matarrese}, \& {Messina}}]{1993MNRAS.260..572C}
{Coles}, P., {Moscardini}, L., {Plionis}, M., {et~al.} 1993, \mnras, 260, 572

\bibitem[{{Coles} \& {Plionis}(1991)}]{1991MNRAS.250...75C}
{Coles}, P., \& {Plionis}, M. 1991, \mnras, 250, 75

\bibitem[{{Colley}(1997)}]{1997ApJ...489..471C}
{Colley}, W.~N. 1997, \apj, 489, 471

\bibitem[{Colley {et~al.}(2000)Colley, Gott, Weinberg, Park, \&
  Berlind}]{Colley:1999rn}
Colley, W.~N., Gott, J.~R., Weinberg, D.~H., Park, C., \& Berlind, A.~A. 2000,
  Astrophys. J., 529, 795

\bibitem[{{Doroshkevich}(1970)}]{1970Ap......6..320D}
{Doroshkevich}, A.~G. 1970, Astrophysics, 6, 320

\bibitem[{Ducout {et~al.}(2013)Ducout, Bouchet, Colombi, Pogosyan, \&
  Prunet}]{Ducout:2012it}
Ducout, A., Bouchet, F., Colombi, S., Pogosyan, D., \& Prunet, S. 2013, Mon.
  Not. Roy. Astron. Soc., 429, 2104

\bibitem[{{Dunkley} {et~al.}(2009){Dunkley}, {Komatsu}, {Nolta}, {Spergel},
  {Larson}, {Hinshaw}, {Page}, {Bennett}, {Gold}, {Jarosik}, {Weiland},
  {Halpern}, {Hill}, {Kogut}, {Limon}, {Meyer}, {Tucker}, {Wollack}, \&
  {Wright}}]{2009ApJS..180..306D}
{Dunkley}, J., {Komatsu}, E., {Nolta}, M.~R., {et~al.} 2009, \apjs, 180, 306

\bibitem[{Eisenstein {et~al.}(2005)}]{Eisenstein:2005su}
Eisenstein, D.~J., {et~al.} 2005, Astrophys. J., 633, 560

\bibitem[{Gay {et~al.}(2012)Gay, Pichon, \& Pogosyan}]{Gay:2011wz}
Gay, C., Pichon, C., \& Pogosyan, D. 2012, Phys. Rev., D85, 023011

\bibitem[{Gorski {et~al.}(2005)Gorski, Hivon, Banday, Wandelt, Hansen,
  Reinecke, \& Bartelman}]{Gorski:2004by}
Gorski, K.~M., Hivon, E., Banday, A.~J., {et~al.} 2005, ApJ., 622, 759

\bibitem[{Gott {et~al.}(2007)Gott, Colley, Park, Park, \&
  Mugnolo}]{Gott:2006za}
Gott, J.~R., Colley, W.~N., Park, C.-G., Park, C., \& Mugnolo, C. 2007, MNRAS,
  377, 1668

\bibitem[{Gott {et~al.}(1986)Gott, Dickinson, \& Melott}]{Gott:1986uz}
Gott, J.~R., Dickinson, M., \& Melott, A.~L. 1986, ApJ., 306, 341

\bibitem[{{Gott} {et~al.}(1992){Gott}, {Mao}, {Park}, \&
  {Lahav}}]{1992ApJ...385...26G}
{Gott}, J.~R., {Mao}, S., {Park}, C., \& {Lahav}, O. 1992, \apj, 385, 26

\bibitem[{{Gott} {et~al.}(1990){Gott}, {Park}, {Juszkiewicz}, {Bies},
  {Bennett}, {Bouchet}, \& {Stebbins}}]{1990ApJ...352....1G}
{Gott}, J.~R., {Park}, C., {Juszkiewicz}, R., {et~al.} 1990, \apj, 352, 1

\bibitem[{{Gott} {et~al.}(1987){Gott}, {Weinberg}, \&
  {Melott}}]{1987ApJ...319....1G}
{Gott}, J.~R., {Weinberg}, D.~H., \& {Melott}, A.~L. 1987, \apj, 319, 1

\bibitem[{{Gott} {et~al.}(1989){Gott}, {Miller}, {Thuan}, {Schneider},
  {Weinberg}, {Gammie}, {Polk}, {Vogeley}, {Jeffrey}, {Bhavsar}, {Melott},
  {Giovanelli}, {Hayes}, {Tully}, \& {Hamilton}}]{1989ApJ...340..625G}
{Gott}, J.~R., {Miller}, J., {Thuan}, T.~X., {et~al.} 1989, \apj, 340, 625

\bibitem[{Hamilton {et~al.}(1986)Hamilton, Gott, \& Weinberg}]{Hamilton:1986}
Hamilton, J. S.~A., Gott, J.~R., \& Weinberg, D. 1986, {\apj}, 309, 1

\bibitem[{Hikage {et~al.}(2006)Hikage, Komatsu, \& Matsubara}]{Hikage:2006fe}
Hikage, C., Komatsu, E., \& Matsubara, T. 2006, ApJ., 653, 11

\bibitem[{Hikage {et~al.}(2002)Hikage, Suto, Kayo, Taruya, Matsubara, Vogeley,
  Hoyle, Gott, \& Brinkmann}]{Hikage:2002ki}
Hikage, C., Suto, Y., Kayo, I., {et~al.} 2002, Publ. Astron. Soc. Jap., 54, 707

\bibitem[{Hong {et~al.}(2016)Hong, Park, \& Kim}]{Hong:2016hsd}
Hong, S.~E., Park, C., \& Kim, J. 2016, Astrophys. J., 823, 103

\bibitem[{{Hoyle} {et~al.}(2002){Hoyle}, {Vogeley}, \&
  {Gott}}]{2002ApJ...570...44H}
{Hoyle}, F., {Vogeley}, M.~S., \& {Gott}, J.~R. 2002, \apj, 570, 44

\bibitem[{Hwang {et~al.}(2016)}]{Hwang:2016yme}
Hwang, H.~S., {et~al.} 2016, Astrophys. J., 818, 173

\bibitem[{James(2012)}]{James:2011wm}
James, J.~B. 2012, ApJ., 751, 40

\bibitem[{Jiang {et~al.}(2008)Jiang, Jing, Faltenbacher, Lin, \&
  Li}]{Jiang:2007xd}
Jiang, C.~Y., Jing, Y.~P., Faltenbacher, A., Lin, W.~P., \& Li, C. 2008,
  Astrophys. J., 675, 1095

\bibitem[{Kim {et~al.}(2015)Kim, Park, L'Huillier, \& Hong}]{Kim:2015yma}
Kim, J., Park, C., L'Huillier, B., \& Hong, S.~E. 2015, JKAS, 48, 213

\bibitem[{Kim {et~al.}(2011)Kim, Park, Rossi, Lee, \& Gott}]{Kim:2011ab}
Kim, J., Park, C., Rossi, G., Lee, S.~M., \& Gott, J.~R. 2011, J. Korean
  Astron. Soc., 44, 217

\bibitem[{Kim {et~al.}(2014)Kim, Choi, Kim, Kim, Lee, Shin, \&
  Kim}]{Kim:2014axe}
Kim, Y.-R., Choi, Y.-Y., Kim, S.~S., {et~al.} 2014, ApJS., 212, 22

\bibitem[{Lewis(2013)}]{Lewis:2013hha}
Lewis, A. 2013, Phys. Rev., D87, 103529

\bibitem[{Lewis \& Bridle(2002)}]{Lewis:2002ah}
Lewis, A., \& Bridle, S. 2002, Phys. Rev., D66, 103511

\bibitem[{Lewis {et~al.}(2000)Lewis, Challinor, \& Lasenby}]{Lewis:1999bs}
Lewis, A., Challinor, A., \& Lasenby, A. 2000, Astrophys. J., 538, 473

\bibitem[{Matsubara(1994)}]{Matsubara:1994we}
Matsubara, T. 1994, arXiv:astro-ph/9501076

\bibitem[{{Matsubara}(1996)}]{1996ApJ...457...13M}
{Matsubara}, T. 1996, \apj, 457, 13

\bibitem[{Matsubara(2000)}]{Matsubara:2000mw}
Matsubara, T. 2000, arXiv:astro-ph/0006269

\bibitem[{Matsubara \& Jain(2001)}]{Matsubara:2000dg}
Matsubara, T., \& Jain, B. 2001, ApJ., 552, L89

\bibitem[{{Melott} {et~al.}(1989){Melott}, {Cohen}, {Hamilton}, {Gott}, \&
  {Weinberg}}]{1989ApJ...345..618M}
{Melott}, A.~L., {Cohen}, A.~P., {Hamilton}, A.~J.~S., {Gott}, J.~R., \&
  {Weinberg}, D.~H. 1989, \apj, 345, 618

\bibitem[{{Melott} {et~al.}(1988){Melott}, {Weinberg}, \&
  {Gott}}]{1988ApJ...328...50M}
{Melott}, A.~L., {Weinberg}, D.~H., \& {Gott}, J.~R. 1988, \apj, 328, 50

\bibitem[{{Park} \& {Gott}(1991)}]{1991ApJ...378..457P}
{Park}, C., \& {Gott}, J.~R. 1991, \apj, 378, 457

\bibitem[{{Park} {et~al.}(2001){Park}, {Gott}, \& {Choi}}]{2001ApJ...553...33P}
{Park}, C., {Gott}, J.~R., \& {Choi}, Y.~J. 2001, \apj, 553, 33

\bibitem[{{Park} {et~al.}(1992){Park}, {Gott}, {Melott}, \&
  {Karachentsev}}]{1992ApJ...387....1P}
{Park}, C., {Gott}, J.~R., {Melott}, A.~L., \& {Karachentsev}, I.~D. 1992,
  \apj, 387, 1

\bibitem[{{Park} {et~al.}(2005){Park}, {Kim}, \& {Gott}}]{2005ApJ...633....1P}
{Park}, C., {Kim}, J., \& {Gott}, J.~R. 2005, \apj, 633, 1

\bibitem[{Park \& Kim(2010)}]{Park:2009ja}
Park, C., \& Kim, Y.-R. 2010, ApJ., 715, L185

\bibitem[{Park {et~al.}(1994)Park, Vogeley, Geller, \& Huchra}]{Park:1994fa}
Park, C., Vogeley, M.~S., Geller, M.~J., \& Huchra, J.~P. 1994, Astrophys. J.,
  431, 569

\bibitem[{Pogosyan {et~al.}(2009)Pogosyan, Gay, \& Pichon}]{Pogosyan:2009rg}
Pogosyan, D., Gay, C., \& Pichon, C. 2009, Phys. Rev., D80, 081301, [Erratum:
  Phys. Rev.D81,129901(2010)]

\bibitem[{Protogeros \& Weinberg(1997)}]{0004-637X-489-2-457}
Protogeros, Z. A.~M., \& Weinberg, D.~H. 1997, The Astrophysical Journal, 489,
  457

\bibitem[{Ryden {et~al.}(1989)Ryden, Melott, Craig, Gott, Weinberg, Scherrer,
  Bhavsar, \& Miller}]{Ryden:1988rk}
Ryden, B.~S., Melott, A.~L., Craig, D.~A., {et~al.} 1989, ApJ., 340, 647

\bibitem[{Schmalzing \& Gorski(1998)}]{Schmalzing:1997uc}
Schmalzing, J., \& Gorski, K.~M. 1998, MNRAS, 297, 355

\bibitem[{Schmalzing {et~al.}(1996)Schmalzing, Kerscher, \&
  Buchert}]{Schmalzing:1995qn}
Schmalzing, J., Kerscher, M., \& Buchert, T. 1996, Proc. Int. Sch. Phys. Fermi,
  132, 281

\bibitem[{Speare {et~al.}(2015)Speare, Gott, Kim, \& Park}]{Speare:2013qma}
Speare, R., Gott, J.~R., Kim, J., \& Park, C. 2015, ApJ., 799, 176

\bibitem[{Wang {et~al.}(2012)Wang, Chen, \& Park}]{Wang:2010ug}
Wang, X., Chen, X., \& Park, C. 2012, ApJ., 747, 48

\bibitem[{{Wang} {et~al.}(2015){Wang}, {Park}, {Xu}, {Chen}, \&
  {Kim}}]{2015ApJ...814....6W}
{Wang}, Y., {Park}, C., {Xu}, Y., {Chen}, X., \& {Kim}, J. 2015, \apj, 814, 6

\bibitem[{Watts {et~al.}(2017)Watts, Elahi, Lewis, \& Power}]{Watts:2017lzm}
Watts, A.~L., Elahi, P.~J., Lewis, G.~F., \& Power, C. 2017, Mon. Not. Roy.
  Astron. Soc., 468, 59

\bibitem[{{Weinberg} {et~al.}(1987){Weinberg}, {Gott}, \&
  {Melott}}]{1987ApJ...321....2W}
{Weinberg}, D.~H., {Gott}, J.~R., \& {Melott}, A.~L. 1987, \apj, 321, 2

\bibitem[{Zunckel {et~al.}(2011)Zunckel, Gott, \& Lunnan}]{Zunckel:2010eh}
Zunckel, C., Gott, J.~R., \& Lunnan, R. 2011, MNRAS, 412, 1402

\end{thebibliography}

\end{document}